\documentclass[a4paper,fleqn,usenatbib]{mnras}

\usepackage{newtxtext,newtxmath}

\usepackage[T1]{fontenc}
\usepackage{ae,aecompl}

\usepackage{graphicx}	
\usepackage{amsmath}	
\usepackage{amssymb}	

\DeclareMathOperator{\mand}{and}
\DeclareMathOperator{\popcnt}{popcnt}

\newcommand{\mpcoh}{\,h^{-1}\,{\rm Mpc}}

\title[Correcting for missing observations]{Unbiased clustering estimation in the presence of missing observations}

\author[Davide Bianchi \& Will J. Percival]{
Davide Bianchi,$^{1}$\thanks{E-mail: davide.bianchi@port.ac.uk}
and Will J. Percival$^{1}$
\\
$^{1}$Institute of Cosmology \& Gravitation, Dennis Sciama Building, University of Portsmouth, Portsmouth, PO1 3FX, UK
}

\date{Accepted XXX. Received YYY; in original form ZZZ}

\pubyear{2017}

\begin{document}
\label{firstpage}
\pagerange{\pageref{firstpage}--\pageref{lastpage}}
\maketitle

\begin{abstract}
  In order to be efficient, spectroscopic galaxy redshift surveys do
  not obtain redshifts for all galaxies in the population
  targeted. The missing galaxies are often clustered, commonly leading
  to a lower proportion of successful observations in dense
  regions. One example is the close-pair issue for SDSS spectroscopic
  galaxy surveys, which have a deficit of pairs of observed galaxies
  with angular separation closer than the hardware limit on placing
  neighbouring fibres. Spatially clustered missing observations will
  exist in the next generations of surveys. Various schemes have
  previously been suggested to mitigate these effects, but none works
  for all situations. We argue that the solution is to link the
  missing galaxies to those observed with statistically equivalent
  clustering properties, and that the best way to do this is to rerun
  the targeting algorithm, varying the angular position of the
  observations. Provided that every pair has a non-zero probability of
  being observed in one realisation of the algorithm, then a
  pair-upweighting scheme linking targets to successful observations,
  can correct these issues. We present such a scheme, and demonstrate
  its validity using realisations of an idealised simple survey
  strategy.
\end{abstract}




\section{Introduction}

The clustering of galaxies observed in spectroscopic galaxy surveys
provides a wealth of cosmological information. In order to extract
this information we need to isolate and remove, or ignore, spatial
galaxy-density fluctuations that arise from non-cosmological sources,
including those that result from the way that observations are
made. One potential source of these fluctuations is that of missing
observations. For surveys using multi-object spectrographs to observe
a target sample of galaxies selected from imaging surveys, it is
usually prohibitively inefficient to observe and obtain spectra for
100\% of the targets. The difficulty results from a combination of the
anisotropic distribution of galaxies on the sky, a product of the very
clustering to be measured, and the mechanical design of the
instrument. Surveys therefore leave a small percentage of the target
sample without spectra. The angular distribution of the missing
galaxies depends on both the observing strategy (for example the
number of times the survey covered a particular region), and the
density of targets, and thus can produce a significant clustering
signal.

For the updated Sloan telescope \citep{gunn2006} as used by the Baryon
Oscillation Spectroscopic Survey \citep[BOSS;][]{dawson2013} , the
fibres cannot be placed closer than 62$\arcsec$ on the focal plane,
and so if two targets are closer than this separation they cannot both
be observed with a single pass of the instrument.  Approximately
$\sim5$\% of the targets in the final Data Release 12 \citep{alam2015} of BOSS were not observed as a consequence
of fibre-collision \citep{reid2016}. Because of the density-dependence
of this sample, fibre-collisions have a strong effect on the
small-scale clustering measurements, as described by \citet{hahn2017},
for example. The scales affected become even larger for deeper surveys
such as the extended BOSS \citep{dawson2016} and the Dark Energy
Spectroscopic Instrument \citep[DESI;][]{amir2016a, amir2016b}.

To construct its main survey covering 14000\,deg$^2$, DESI will make
approximately 10,000 observations, taking 5000 spectra in each 7.5\,deg$^2$ field-of-view.
Although the average number of
observations covering any patch in the survey is 5, the range is
between 1 and 12. In regions of high target density, and for targets
of low priority in the ranking of different target classes, there will
be missing observations. Thus, unless corrected they have the
potential to significantly distort measurements of cosmological
clustering \citep{pinol2017,burden2017}.

In this paper, we consider the general problem of missing galaxies, in
a way that is not tied to any survey, and present an algorithm for
debiasing the measured correlation function. It works by determining a
probability of selection for any pair and weighting by the inverse of
this probability. This then provides an unbiased estimation of the
correlation function, provided that any pair in the sample has a
non-zero probability of being observed if it were moved to some
location in the survey. The layout of our paper is as follows: in
Sec.~\ref{sec miss} we review the problem of missing observations; in
Sec.~\ref{sec gen idea} we present the derivation of our estimator; in
Sec.~\ref{sec mr} we define the selection algorithm that we use for
testing; in Sec.~\ref{sec galaxy} we discuss the behaviour of (a
simplified version of) the estimator compared to that of the nearest
neighbour assignment\footnote{This discussion is, to some extent, pedagogical, the reader interested in a compact description of the full estimator might want to skip this section at first reading.}; in Sec.~\ref{sec pair weights} we present the
practical implementation of the estimator, which we compare to
simulations in Sec.~\ref{sec simulations}; we conclude summarising our
results in Sec.~\ref{sec conclusions}.

\section{Missing spectroscopic observations}\label{sec miss}

We consider a general redshift survey, consisting of a set of targets
with known angular positions, that we want to spectroscopically
observe.  If a randomly selected sample of targets does not have
spectroscopic observations, then our estimate of the 3-dimensional
overdensity at any location from the observed sample is unbiased,
provided that the expected number of observations is
reduced. e.g. suppose we define
\begin{equation}
  \delta({\bf x}) = \frac{\rho_{\rm all}({\bf x})}{\langle\rho_{\rm all}\rangle}-1,
\end{equation}
then this $\delta$ is unaltered by the transformation $\rho_{\rm all}({\bf x})
\to \alpha\rho_{\rm all}({\bf x})$ for any $\alpha<1$ that is
spatially invariant. 

We also do not need to worry about missing redshift measurements as a
function of galaxy type. For example, suppose we target two classes of
galaxies, each with a linear deterministic bias
$\delta_{{\rm gal},A}=b_A\delta_{\rm mass}$,
$\delta_{{\rm gal},B}=b_B\delta_{\rm mass}$, but only measure
redshifts for galaxies in class $A$. Provided that we use
$\langle\rho_A\rangle$ in the denominator when calculating $\delta$,
then our estimate of $\delta$ only depends on the observed galaxies,
and is unaffected by sample $B$.

Many surveys are not able to spectroscopically observe the full target
sample, and make observations based on the angular target density. For
example, the multi-object spectrograph on the Sloan telescope
\citep{gunn2006} cannot simultaneously observe two targets closer than
$62$\,\arcsec. This leads to a deficit of small angular separation
pairs of galaxies, which is particularly severe for regions of the sky
covered by only one pass of the instrument.  In order to correct these
effects, a number of approximate methods have been put forward
\citep{anderson2012, guo2012, hahn2017}.  The standard approach
adopted by the BOSS team has been to upweight by one the nearest
target to each missing target \citep{anderson2012}. To see how this
works, consider pairs of galaxies as counted in standard correlation
function measurements: the target nearest to that missed is
statistically identical as there was a 50/50 chance as to which was
observed, and it consequently has the same expected clustering
properties. The upweight therefore approximately corrects the total
pair count for missed pairs between the missed targets and other
targets outside of the pair in question. The pair between the missed
and nearest target is still excluded, and leads to a small-scale bias.
\citet{guo2012} suggested an algorithm that uses the regions of
overlapping observations to understand those missed.  However it does
not works perfectly, because, as we discuss later, the observed pairs
are not statistically identical to those missed.
\citet{reid2014} adopted a different approach where they assigned each missing galaxy the redshift of the nearest observed galaxy.
This artificially creates small-separation pairs, but not necessarily with the correct distribution.

The situation is likely to be significantly worse for future surveys
such as DESI \citep{amir2016a, amir2016b}, which will make observations using a
grid of fibre feeds, with each fibre able to move independently, but
only within its patrol radius.  Even though the targets will be
observed with multiple passes, the final set of spectroscopically
observed targets will exhibit strong angular-density dependence.  Two
recent papers presented methods to combat the effect of missing
galaxies due to the fibre assignment scheme of DESI.
\citet{pinol2017} showed that allowing for variations in coverage
within the mask, commonly quantified by a random Poisson sampling and
referred to as the ``random catalogue'' reduces this effect.  They
argue that the best way to completely mitigate the effect is to remove
the angular modes from the analysis.  \citet{burden2017} advocate a
similar approach, modifying the standard correlation function
estimator in order to null angular modes, and demonstrated how this
would work using mock data. Note that both of these approaches discard
information rather than trying to understand and model the effects.

Given that we know the angular distribution of the targets, it has
been suggested that, when calculating the 3-dimensional correlation
function, we upweight each observed pair by the reciprocal of the
fraction of observed pairs of targets with that angular separation
\citep{hawkins2003}. This correction does not work in general,
because, again, it assumes that the radial properties of the
unobserved pairs of targets are statistically equivalent to those of
the observed pairs. Let us consider the example of the SDSS, given
above. Here, missing close-pairs are more likely to be in triplets of
targets than observed close-pairs: triples require three observations
to fully observe, whereas doubles only require two, and the area
covered by three observations is significantly smaller than that
covered by two. Galaxies in triples of targets are more likely to be
radially associated than galaxies in doubles, as they represent more
unlikely chance alignments.  The idea of upweighting of angular pairs
is similar to the method put forward by \citet{hahn2017}, who
probabilistically assigned galaxy redshifts to missing galaxies based
on those observed. Both approaches use the observed galaxies to
understand the unobserved ones, but the problem is also the same as
that discussed above - that the missing pairs or galaxies and observed
pairs or galaxies need to be carefully matched: the matching between
observed and unobserved pairs is at the heart of any scheme to correct
for missing observations.

\section{The new algorithm}\label{sec gen idea}

In this section we present a new method to match observed and
unobserved pairs. As we consider pairs of galaxies, it is easiest to
consider this in the context of the measurement of the correlation
function. We argue that this matching between observed and unobserved
targets is simpler if we work with pairs rather than galaxies, as the
selection algorithm can act over large scales, meaning it is difficult
to select galaxies in the observed sample that match those that are
missing. Because the calculation of the correlation function only
depends on the numbers of pairs, by matching pairs we can be sure that
we are including all of the necessary information.

One final assumption we make is that all pairs have non-zero probability that they could be observed were we allowed the freedom to move them to any spatial location covered by the observations.
So there are no pairs of targets that represent objects that
could never be observed. Both the SDSS BOSS and eBOSS surveys and DESI
match this requirement.

\subsection{The effect of adding and removing galaxies}

We consider a given realisation of some anisotropically clustered
random field, traced by a set of particles.  We can measure the number
of pairs in a given separation bin $\vec{s} \pm \Delta \vec{s}/2$, which
we refer to as $DD(\vec{s})$.  Suppose that we choose one galaxy and
remove from our counts all the pairs formed by this galaxy, but we
count twice the pairs formed by another galaxy.  Plus we include in
the counts the single pair formed by these two galaxies.  We then have
a new value $DD_1(\vec{s}) \neq DD(\vec{s})$.  Similarly, we can
interchange the two selected galaxies and get
$DD_2(\vec{s}) \neq DD_1(\vec{s}) \neq DD(\vec{s})$.  Trivially,
$[DD_2(\vec{s}) + DD_1(\vec{s})] / 2 = DD(\vec{s})$, or, in other
words, the mean of the two new counts corresponds to the original one.
If realisation 1 and 2 are statistically equivalent, i.e. the
probability of having 1 or 2 is apriori identical, their mean
corresponds to the expected value of an unbiased estimator for
$DD(\vec{s})$.  This simple argument can be invoked to justify
standard countermeasures against the fibre-collision issue, such as
the nearest neighbour upweighting \citep[e.g.][]{anderson2012}.  We
will show that this class of weighting schemes can be seen as
approximations, formally not unbiased, of a more rigorous and general
description of the problem.

\subsection{Unbiased estimator}\label{sec unbiased}

The evaluation of two-point statistics in a galaxy survey is based on pair
counts at different separations $\vec{s}$, e.g. if we want to measure
the correlation function a standard approach is to use the following
estimator \citep{landy1993}:
\begin{equation}
\hat{\xi}(\vec{s}) = \frac{DD(\vec{s})}{RR(\vec{s})} - 2 \frac{DR({\vec{s}})}{RR(\vec{s})} + 1 \ ,
\end{equation}
where $DD$ is the number of data (i.e. galaxy) pairs, $RR$ is the
number of pairs in a random catalogue covering the same volume of the
survey and $DR$ is the number of data-random pairs.


Suppose we have an algorithm to extract a subset of galaxies from the
full sample according to some arbitrary selection rule.  Since this
selection algorithm is completely free, in general, the pair counts
$DD(\vec{s})$ in the new sample and those from the original parent
sample $DD_p(\vec{s})$ will differ, in both shape and amplitude (and
similarly for $DR$).  Suppose that, as in any realistic scenario, the
algorithm is stochastic, i.e. for a given galaxy sample there are many
possible outcome subsets, corresponding to different random seeds.
The quantity of interest is then the expectation value of $DD$ and
$DR$ (obviously $RR=RR_p$ remains unchanged).  Still, for a generic
algorithm $\langle DD \rangle \neq DD_p$, and similarly for $DR$.

If we denote with $p_i$ the probability of the $i$-th galaxy of being
selected, the probability that the pair formed by the $m$-th and
$n$-th galaxies contributes to the counts is
\begin{equation}\label{eq pmn}
p_{mn} = p_m p_n (1 + c_{mn} )  \ ,
\end{equation}
where
\begin{equation}\label{eq c}
c_{mn} \equiv \ \frac{p_{mn}}{p_m p_n} - 1
\end{equation}
is the selection correlation associated to that specific pair.

Each pair carries two fundamental pieces of information: the
separation $\vec{x}_m - \vec{x}_n$ and the selection probability
$p_{mn}$.  As we will see, in general the latter cannot be reduced to
the former.  It is natural to use this probability to correct the
galaxy pair counts.  Specifically, we define the statistical weight of
each pair as
\begin{equation}
w_{mn} \equiv \frac{1}{p_{mn}} \ .
\end{equation} 
At any separation the pair count is then given by 
\begin{equation}\label{eq pw}
DD(\vec{s}) = \sum_{\vec{x}_m - \vec{x}_n \approx \vec{s}} w_{mn} \ ,
\end{equation}
where the symbol ``$\approx$'' means that the sum is performed over
pairs whose separation falls in a specific $\vec{s}$ bin.  Obviously
only pairs selected by the algorithm are considered\footnote{ With a
  more rigorous notation,
  $DD(\vec{s}) =\frac{1}{2} \sum_{m \ne n} w_{mn} \ l_m \ l_n \
  l_{mn}(\vec{s})$
  where $l_i$ is a logical weight such that $l_i = 1$ if the $i$-th
  galaxy has been selected and $l_i = 0$ otherwise.  Similarly,
  $l_{mn} = 1$ if the pair belongs to the specific separation bin
  under exam and $l_{mn} = 0$ otherwise (obviously $w_{mn} = w_{nm}$
  and $l_{mn} = l_{nm}$).}.

By construction, {\it if each pair has non-zero probability of being
  selected}, the expectation value of the so obtained $DD$ is
unbiased, i.e. $\langle DD \rangle = DD_p$.  This can be understood by
observing that, with the pairwise-inverse-probability (PIP) weighting
scheme just introduced, if we sum over $N$ realisations,
statistically, each pair appears $N$ times and, as a consequence, each
pair contributes a term $N/N=1$ to the average pair counts.  On scales
where at least one of the pairs has null selection probability, the PIP
scheme is potentially biased, reflecting the fact that the information
on that pair is completely lost.  Trivially, when a pure
fibre-collision issue is considered, no pair below some minimum-fibre-separation scale $r_f$ can be observed\footnote{For the sake of simplicity, we can think of $r_f$ as the fibre diameter.} and the estimator is not only biased but completely uninformative on such scales.

Inspired by Eq. (\ref{eq pmn}), we can rearrange the pair weights as follows
\begin{equation}
w_{mn} = w_m \ w_n \ w^{(c)}_{mn} \ ,
\end{equation}
where we defined $w_i \equiv 1/p_i$ and
$w^{(c)}_{mn} \equiv 1/(1+c_{mn})$.  This makes clear that, {\it if
  the selection correlation is negligible}, pair weighting can be
reduced to galaxy weighting, i.e. $w_{mn} = w_m w_n$.

As regards $DR$ counts, galaxy weighting is always sufficient, since
the selection algorithm does not apply to the random sample and, as a
consequence, the selection probability of a galaxy-random pair always
reduces to the individual probability of the galaxy.

Note that all the above considerations do not necessarily have be
related to a fibre-collision issue.  The description is formally valid
for any scenario in which a subset of particles is extracted from a
larger sample with known selection probability.

\section{Selection algorithm and correlation length}\label{sec mr}

In the following we focus more explicitly on a fibre-collision-like
problem, which means we consider only selection criteria based on the
angular position of the galaxies.  For simplicity, we assume the plane
parallel approximation, i.e. the angular separation of pairs
corresponds to the perpendicular-to-the-line-of-sight separation
$s_\perp$.

In order to help to demonstrate the more general idea, we define a
specific selection algorithm, which we use for testing.  We explicitly
discuss throughout the paper which of our results depends on this
particular choice.  The algorithm we adopt is meant to maximise the
randomness of the selection criteria in the presence of fibre
collisions, which is why hereafter we refer to it as the maximum
randomness (MR) algorithm.  It can be summarised as follows: we
randomly pick a pair among those with angular separation smaller than
$r_f$ and randomly discard one of the two galaxies; we iterate the
procedure until there are no more pairs with angular separation
smaller than $r_f$.

Given the geometry of problem we are studying, it is useful to
introduce the concept of angular friend-of-friend (AFOF) halo,
obtained by restricting the standard friend-of-friend definition
\citep[e.g.][]{davis1985} to the angular separation only,
i.e. ignoring the line-of-sight coordinates of the galaxies, with
linking length given by the minimum-fibre-separation scale~$r_f$.

With the MR algorithm the selection probability of individual galaxies
is independent on scales larger than $r_c$, this latter
being the largest separation between two galaxies belonging to the
same AFOF halo, or, roughly speaking, the size of the largest AFOF
halo in the sample.  In other words, whether a galaxy is selected or
not in general depends on all the other galaxies belonging to the same
halo, but not on the galaxies outside that specific halo.
For more complex algorithms we can think of generalising $r_c$ as the
correlation length above which the selection correlation introduced in
Eq.~(\ref{eq c}) becomes negligible.  In any case, for $s_\perp > r_c$
the pairwise probability reduces to the product of individual
probabilities, and our unbiased estimator, Eq. \ref{eq pw}, can be
expressed in terms of galaxy weights,
\begin{equation}\label{eq iip}
DD(\vec{s}) = \sum_{\vec{x}_m - \vec{x}_n \approx \vec{s}} w_iw_j \ .
\end{equation}
These individual-inverse-probability (IIP) weights can be evaluated
analytically if the selection algorithm is simple enough (see
Sec.~\ref{sec case study}) or, more realistically, estimated
numerically.  Note that $r_c$ is a well-defined number that can be
measured from the sample under examination.  For $s_\perp < r_c$ we
need to enforce pair weighting as specified in Eq.~(\ref{eq pw}).
Unfortunately, $r_c$ grows fast with the galaxy number density and the
collision scale $r_f$, thus making pair weighting in general
preferable.  Analogously to the individual one, the pair probability
can be in principle computed analytically or, more pragmatically,
evaluated numerically, with the obvious complication of having to deal
with $N^2$ objects rather than just $N$.

\section{Galaxy weighting}\label{sec galaxy}

In this section we compare two examples of individual-galaxy weighting
schemes, namely the IIP approach, defined by Eq.~(\ref{eq iip}), and
the well know nearest neighbour (NN) correction, which consists of
assigning the weight of the missing galaxy to its nearest (in terms of
angular position) observed companion.  Two more galaxy-weight
prescriptions, with performances comparable to those of the NN
assignment, are considered in App.~\ref{sec NNvar}.  As discussed
above, weighting individual galaxies is not the most general possible
approach to the problem of missing observation, since it does not
account for selection correlation.  Dealing with this issue actually
requires a pair-weighting approach, which we present in Sec.~\ref{sec
  pair weights}.  It is nonetheless instructive to see how, even in
this simplified scenario, a probability-oriented reasoning is
convenient with respect to the more standard idea of moving weights
from the missing to the observed galaxies, which is behind the NN
correction.

\subsection{Case study}\label{sec case study}

Here we discuss a simple example of a small structure of target
galaxies, which hopefully will help to clarify a few basic concepts.
We consider a single AFOF structure, sketched in Fig.~\ref{fig
  triplet}.
\begin{figure}
 \begin{center}
  \vspace{-1.5cm}
   \includegraphics[width=9cm]{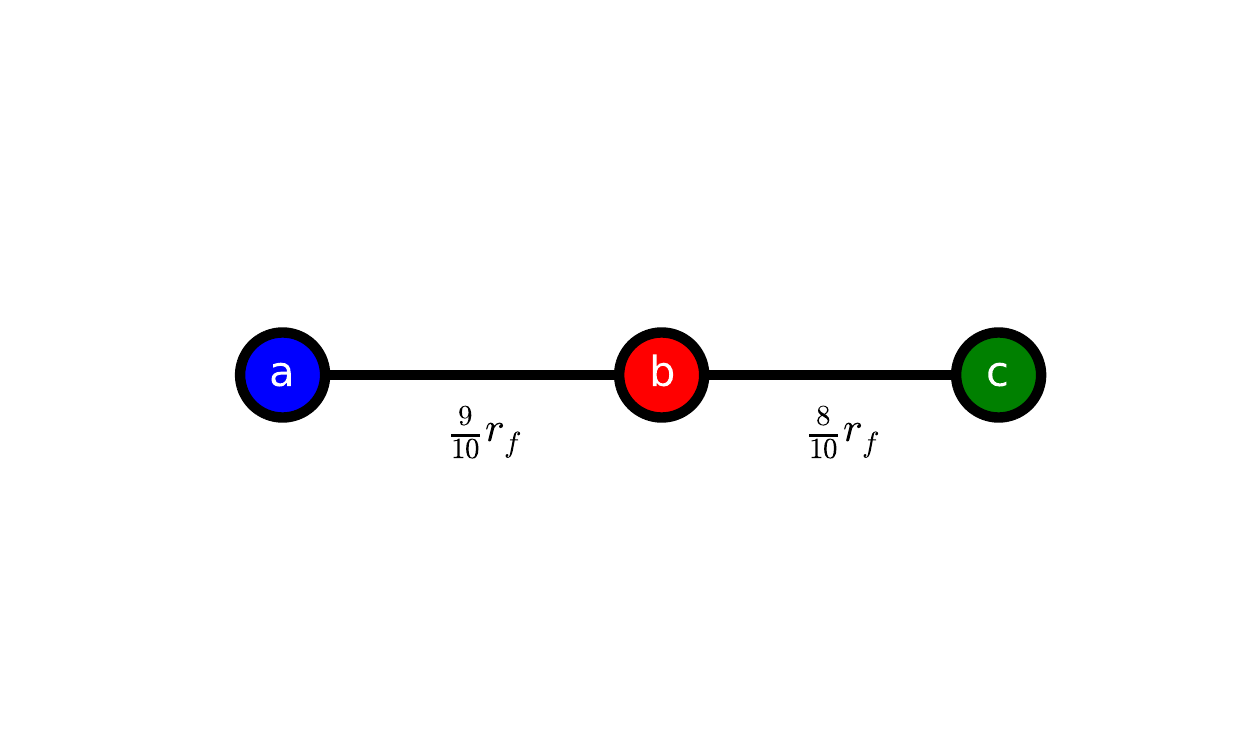}\\
   \vspace{-1.5cm}
   \caption{Partially collided triplet formed by the collided pairs
     $\{a,b\}$ and $\{b,c\}$. The line of sight is perpendicular to
     the plane of the figure. The intra-particle separations are
     expressed in units of the collision length $r_f$, which in a pure
     fibre-collision problem is just the size of the fibre. Only
     collided pairs, i.e. those with separations smaller than $r_f$
     are connected by solid lines.}
   \label{fig triplet}
 \end{center}
\end{figure}
We define the triplet $\{a,b,c\}$ as a {\it partially collided}
structure, formed by two {\it collided} structures, the pairs
$\{a,b\}$ and $\{b,c\}$.  We consider pairs with different separation
just to avoid degeneracy when applying the NN scheme.  All the
calculations in this section refer to the MR algorithm defined in
Sec.~\ref{sec mr}.  When applied to the AFOF halo in the figure, the
NN scheme yields
\begin{equation}
S^{(NN)} =
\begin{bmatrix}
    1 & 0 & 2 \\
    0 & 3 & 0 \\
    0 & 0 & 3 \\
    3 & 0 & 0 
\end{bmatrix} \quad , \quad
P^{(NN)} =
\begin{bmatrix}
    1/2  \\
    1/4  \\
    1/8 \\
    1/8 
\end{bmatrix} \ , 
\end{equation}
where each row of the matrix $S^{(NN)}$ represents one of the possible
set of weights $\{w_a, w_b, w_c\}$ associated to the galaxy triplet.
The array $P^{(NN)}$ represents the correspondent probability, which
is evaluated analytically.  Trivially, only selected galaxies can have
non-zero weight.  The number of objects is conserved, i.e. the sum of
the elements of a row is always 3.  The sum of the elements of the
columns, weighted by the correspondent probability, is
$\{11/8, 3/4, 7/8\}$.  This means that the estimator is biased, since,
in order to play the game described in Sec. \ref{sec gen idea} we need
this sum to be $\{1, 1, 1\}$.  More explicitly, if the AFOF halo under
examination is the only collided structure in the universe, then the
selection probability of the cross pairs formed by a galaxy belonging
to the halo with all the external ones is exactly given by the
individual probability of the former.  As a consequence, when the
weighted sum is $\{1, 1, 1\}$ the cross-pair count is formally
unbiased, in the sense that its mean is exactly what we would have
without any selection process (i.e. fibre collision).  For this
specific example, it means $\langle DD \rangle = DD_p$ on scales
$s_\perp > 17/10 \ r_f$, which is the size of the largest pair in the
halo, namely pair $\{a,c\}$.  This reasoning can easily be extended to
the general scenario in which there are several AFOF halos in the
sample, since the resulting cross probabilities are disjoint by
construction\footnote{\label{note NN}For simplicity, we assume that
  the weight of the missing galaxies is always transferred to galaxies
  belonging to the same AFOF halo, which is not necessarily the case
  when NN assignment is coupled to the MR algorithm.}.

With the IIP scheme we instead obtain
\begin{equation}
S^{(IIP)} =
\begin{bmatrix}
    8/5 & 0 & 8/5 \\
    0 & 4 & 0 \\
    0 & 0 & 8/5 \\
    8/5 & 0 & 0 
\end{bmatrix} \quad , \quad
P^{(IIP)} =
\begin{bmatrix}
    1/2  \\
    1/4  \\
    1/8 \\
    1/8 
\end{bmatrix} \ . 
\end{equation}
In this case each galaxy is weighted by its inverse probability of
being selected by the algorithm.  At variance with $S^{(NN)}$, for
$S^{(IIP)}$ the number of objects is not conserved, but the estimator
is unbiased, since the weighted sum of the elements of each column is
$\{1, 1, 1\}$ by construction.  The fact that the galaxy number is not
conserved suggests that the higher accuracy of this estimator comes at
the cost of less precision (i.e. no bias but larger variance).  We
show in Sec.~\ref{sec var} how to circumvent this issue.

One interesting question is whether the pair counts are correct inside
the structure under consideration.  Obviously, this cannot be the case
for pairs with angular separation $s_\perp < r_f$ since none of these
pairs can be observed by definition, but it could still be true for
the pair $\{a,c\}$.  Indeed, for the NN scheme, the pair $\{a,c\}$ is
correctly weighted, in the sense that, when summing over $N$ different
realisations, statistically, this pair is counted $N$ times, i.e one
time in average: $\sum_i S^{(NN)}_{i1} S^{(NN)}_{i3} P^{(NN)}_i = 1$.
One might wonder if this is a general property of the NN assignment.
It is then useful to consider a further example, which shows that this
is not the case.
\begin{figure}
 \begin{center}
   \vspace{-0.4cm}
   \includegraphics[width=9cm]{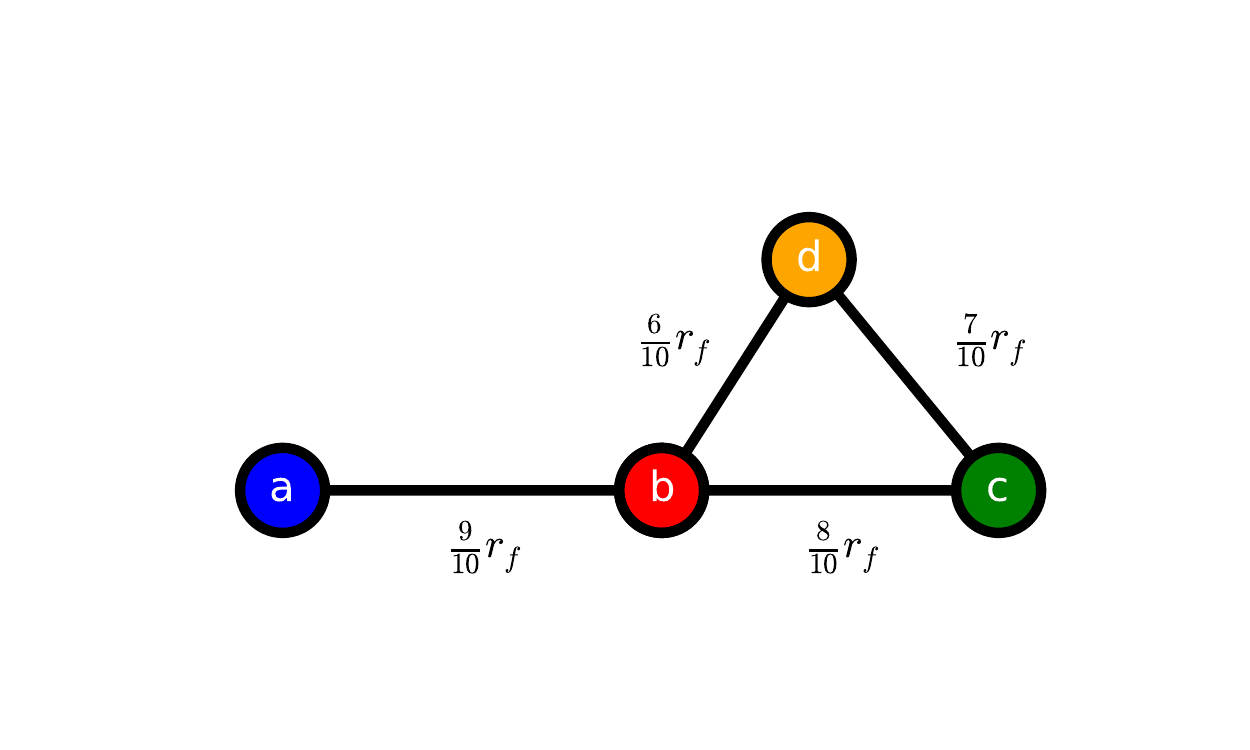}\\
   \vspace{-0.7cm}
   \caption{Partially collided structure formed by the collided pair
     $\{a,b\}$ and the collided triplet $\{b,c,d\}$. Same notation as
     in Fig.~\ref{fig triplet}.}
   \label{fig triplet and pair}
 \end{center}
\end{figure}   
We repeat our calculations for a new AFOF structure, obtained by
adding a forth galaxy $d$ to the previously discussed triplet, as
sketched in Fig. \ref{fig triplet and pair}.  We get
\begin{equation}
S^{(NN)} =
\begin{bmatrix}
    1 & 0 & 3 & 0 \\
    1 & 0 & 0 & 3 \\
    0 & 4 & 0 & 0 \\
    0 & 0 & 4 & 0 \\
    0 & 0 & 0 & 4 \\
    4 & 0 & 0 & 0 
\end{bmatrix} \quad , \quad
P^{(NN)} =
\begin{bmatrix}
    5/16 \\
    5/16 \\
    1/6 \\
    7/96 \\
    7/96 \\
    1/16 
\end{bmatrix} \ ; 
\end{equation}

 \begin{equation}
S^{(IIP)} =
\begin{bmatrix}
    16/11 & 0 & 96/37 & 0 \\
    16/11 & 0 & 0 & 96/37 \\
    0 & 6 & 0 & 0 \\
    0 & 0 & 96/37 & 0 \\
    0 & 0 & 0 & 96/37 \\
    16/11 & 0 & 0 & 0 
\end{bmatrix} \quad , \quad
P^{(IIP)} =
\begin{bmatrix}
    5/16 \\
    5/16 \\
    1/6 \\
    7/96 \\
    7/96 \\
    1/16 
\end{bmatrix} \ . 
\end{equation}
Clearly, the complexity of the analytical calculations grows fast with
the number of particles involved.  As anticipated, the number of pairs
with $s_\perp > r_f$ inside the AFOF halo is in general not conserved
(for $s_\perp < r_f$ this number has to be zero by construction).  For
instance, the average counting of pair $\{a,c\}$ is $15/16$ and
$480/407$ for NN and IIP, respectively, meaning that neither of the two
corrections is unbiased for $r_f < s_\perp < r_c$.  At least for IIP,
this is not a surprise, since on scales smaller than $r_c$, by
definition, the selection correlation cannot be neglected when
evaluating the pairwise probability.

Finally, the comparison between the two AFOF structures,
Figs. \ref{fig triplet} and \ref{fig triplet and pair}, provides us
with the proof that the selection probability cannot be deduced by the
separation only: despite the separation between galaxy $a$ and $c$
being fixed, the selection probability of pair $\{a,c\}$ drops from
$1/2$ to $5/16$ when galaxy $d$ is added.

\subsection{Range of validity}\label{sec range}

In order to summarise the properties of the above estimators, it is
useful to divide the $s_\parallel$-$s_\perp$ plane into three regions.
These regions are defined by the two characteristic scales already
introduced, $r_f$ and $r_c$.  The former represents the minimum allowed
angular separation between galaxies, e.g. the size of the fibre.  The
latter is the largest angular separation between two galaxies
belonging to the same AFOF structure, where the linking length is
$r_f$ (or, in a more general scenario, just a selection-correlation
length). Both the estimators discussed above are potentially biased
for $r_f < s_\perp < r_c$ and completely uninformative for
$s_\perp < r_f$.  The IIP estimator is rigorously unbiased in the
plane $s_\perp > r_c$, whereas the NN assignment is not, unless all
the AFOF halos are purely collided structures (i.e. not partially
collided).  In other words, if we pick a single random galaxy from
each AFOF halo, these latter estimator is unbiased as well\footnote{At
  least under the simplifying assumption discussed in note \ref{note
    NN}.}.  In general, we can think of the NN and similar schemes,
such as those discussed in App. \ref{sec NNvar}, or found in the
literature, e.g. local density weighting \citep[e.g.][]{pezzotta2016},
as an approximation of the IIP scheme.  How reliable these
approximations are strongly depends on the characteristics of the
galaxy survey, such as the fraction of collided pairs among the total
number of partially-collided structures.

\section{Pair weighting}\label{sec pair weights}

So far we have shown that it is convenient to: (i) see weights as
inverse probabilities; (ii) weight pairs rather than individual
galaxies.  In the following we show that this is not only convenient
but also feasible, by providing a practical implementation of the PIP
method, which includes important considerations about how to reduce
the variance of our estimator.  But first we focus on how to extend
our clustering estimate down to arbitrary small separations.

\subsection{Including small scales}\label{sec small scales}

As we have already discussed, the PIP weighting scheme is unbiased by
construction on scales larger than $r_f$, for a pure fibre collision
issue, where $r_f$ is the diameter of the fibre.  For a completely
general selection algorithm, PIP is unbiased on all the scales for
which no pair has null selection probability.  This suggests that an
all-scale unbiased estimator of the two-point functions can be obtained by
removing the $r_f$ constraint in small {\it random} regions of the
survey.  Practically, this can be obtained by observing more than once
a subset, not necessarily connected, of the whole sample.  Overlap
regions in the observing strategy are indeed quite common in modern
surveys.  It is important to emphasise that a 100\% coverage of this
subset is not necessarily needed.  Indeed, in order to break the $r_f$
constraint it suffices to observe twice a region randomly picked from
the total survey area, regardless of the fact that we might still
miss objects in such a region (but, obviously, the more the galaxies
we observe, the more the information that we can extract).

\subsection{Minimising the variance: angular upweighting}\label{sec var}

So far we have focused on the bias of the estimator.  We now move our
attention to the issue of minimising its variance.  To this purpose,
we note that there is further information available, which we have not
used yet, namely the knowledge of the angular correlation function of
the full parent sample.  In a companion paper \citep{percival2017} we
explicitly discuss how, under quite general assumptions, this
information can be used to build minimum variance estimators.
Specifically we show that applying an angular upweighting \citep[AUW,
e.g.][]{hawkins2003} correction to an unbiased estimator whose
variance is nearly Poissonian, has the beneficial effect of minimising
the variance of this latter while leaving its expectation value
unchanged.  We therefore define our final weighting scheme as
\begin{equation}\label{eq PIPauw}
DD(\vec{s}) = \sum_{\vec{x}_m - \vec{x}_n \approx \vec{s}} w_{mn} \frac{DD^{(p)}_a(s_\perp)}{DD_a(s_\perp)} \ ,
\end{equation}
where $DD^{(p)}_a$ and $DD_a$ represent the angular pair counts of the
parent and the observed sample, respectively, whereas $w_{mn}$ is the
PIP factor previously introduced.  Note that $DD_a$ is, in turn,
computed via the same $w_{mn}$ weights.
Analoguosly, for the cross count we define $DR = \sum w_{m} \ DR^{(p)}_a / DR_a$, where the PIP weight of the galaxy-random pair is fully characterised by the individual weight $w_m$ of the galaxy.

Note also that it may be possible to further reduce the variance by a
sensible selection of target galaxies. Any population or
sub-population where only a small fraction of pairs will be recovered
will in general, when added to the full sample, increase the shot
noise of the population as a whole as we will be upweighting a small
number of pairs. By judging the relative shot noise of
sub-populations, if they can be selected from the full target sample,
we should be able to judge whether or not they are worth including in
the analysis.

\subsection{Practical implementation: bitwise weights}\label{sec implement}

In general, finding an efficient PIP-weighting implementation is not a
trivial task.  First, although in principle it is formally possible to
compute analytically the weight associated to each pair, in practice
this requires us to identify all the classes of collided structures in
the sample and solve explicitly for the probability of the pairs there
within, similarly to the calculation presented for the two simple
examples discussed in Sec.~\ref{sec case study}.  Even in the presence
of a very simple selection algorithm, a dense sample is enough to make
the analytical calculations unfeasible, due the complexity of the
resulting collided structures.  This problem can be circumvented by
estimating the probabilities numerically by randomly repeating the
selection process several times and estimating the probability of a
given pair from the frequency with which it is chosen.  Second, future
galaxy surveys will collect spectra from $N_{gal} \sim 10^7$ galaxies,
which implies $N_{gal}^2 \sim 10^{14}$ pair weights with a consequent
storage issue (which becomes catastrophic for higher order
statistics).  

We therefore introduce an effective scheme that retains all of the
information about the pair weighting resulting from repeated
applications of the targeting algorithm, but scales as $N_{gal}$.  The
selection probability of any galaxy can be recorded using
$N_n/N_{bits}$, where $N_n$ is the number of times the $n$-th galaxy
has been chosen and $N_{bits}$ is the total number of realisations of
the selection process.  In other words,
$N_n = \sum_{i=1}^{N_{bits}} s^{(n)}_i$, where ${\bf s}^{(n)}$ is a
logical array of length $N_{bits}$ whose $i$-th element is equal to 1
or 0 according to whether the galaxy has been selected or not in the
$i$-th realisation, e.g.  ${\bf s}^{(n)} = \{1,0,1,\dots,0,0,1\}$.
From these data, the probability associated to a given pair is
$N_{nm}/N_{bits}$, with
$N_{nm} = \sum_{i=1}^{N_{bits}} s^{(n)}_i s^{(m)}_i$. As any logical
array ${\bf s}$ can be seen as the binary representation of an integer
number, the concept of weighting individual objects
rather than pairs can be formally saved using such an approach.

In this case, we should drop the usual idea that the pair weight
should be obtained as the product of galaxy weights, $w_{mn}=w_m w_n$.
Indeed, by defining $w^{(b)}$ ($b$ stands for bitwise) as the integer
number corresponding to ${\bf s}$, we have
\begin{equation}\label{eq wb}
w_{mn} = \frac{N_{bits}}{\popcnt\left[w^{(b)}_m \ \mand \ w^{(b)}_n \right]}
\end{equation}
where $\mand$ and $\popcnt$ are standard (and extremely fast) bitwise
operators.  The former multiplies two integers bit by bit, whereas the
latter is a population-count operator, which takes an integer and
returns the sums of its bits.  Since current computers rely on 32 or
64 bit architectures, for realistic choices of $N_{bits}$, large
enough for an accurate sampling of the selection probability,
e.g. $N_{bits} \sim10^3$, we need to split $w^{(b)}$ into $\sim 10$
sub-weights.  This obviously makes the evaluation of $w_{mn}$ slower
but still tractable\footnote{Note that it is in general convenient to
  adopt an algorithm that first evaluates the IIP weight as
  $w_n = N_{bits} / \sum_{i=1}^{N_{bits}} s^{(n)}_i$ for
  $n=1,\dots,N_{gal}$ and then, when computing $DD$, enforces
  Eq.~(\ref{eq wb}) only if the IIP weight of both galaxies under
  examination is larger than 1, while using $w_{mn}=w_m w_n$
  elsewhere. Also note that the evaluation of the cross-pair counts
  $DR$ only requires IIP weights, i.e. it is not slower than when any
  other standard weighting technique is adopted.  As a consequence,
  PIP weighting becomes more time consuming than other more standard
  schemes only if the cpu time required to evaluate $DD$ becomes
  larger than that required for $DR$.  Since the random catalogue is
  normally at least one order of magnitude denser than the galaxy
  catalogue, this bottleneck issue arises only if we adopt a very
  large number of bits $N_{bits}$.}.  Finally it is important to note
that Eq. (\ref{eq wb}) can be trivially extended to any higher order
statistics just by iterating the $\mand$ operator.

\section{Comparison to simulations}\label{sec simulations}

\begin{table}
 \centering
  \begin{tabular}{rlllllll}
  observing strategy & $\bar{f}_s$ & $\bar{f}_d$ & $f_c$ & $f_u$ & $f_{px}$ & $f_{p1}$\\
  \hline
  \hline
  OS1 & $0.56$ & $0.44$ & $0.75$ & $0.25$ & $0.75$ & $0.25$\\
  OS2 & $0.84$ & $0.16$ & $0.75$ & $0.25$ & $0.49$ & $0.51$\\
  OS2sub & $0.63$ & $0.37$ & $0.75$ & $0.25$ & $0.75$ & $0.25$\\
  OSmulti & $0.66$ & $0.34$ & $0.75$ & $0.25$ & $0.75$ & $0.25$\\
  \end{tabular}
  \caption{From left to right: average fraction of selected galaxies;
    average fraction of discarded galaxies; fraction of collided
    galaxies; fraction of uncollided galaxies; fraction of galaxies
    with selection probability $0<p<1$; fraction of galaxies with
    selection probability $p=1$. Each row corresponds to one of the
    four different observing strategies discussed in Sec.~\ref{sec
      simulations}.}
 \label{tab stats}
\end{table}
\begin{figure*}
 \begin{center}
   \includegraphics[width=8.5cm]{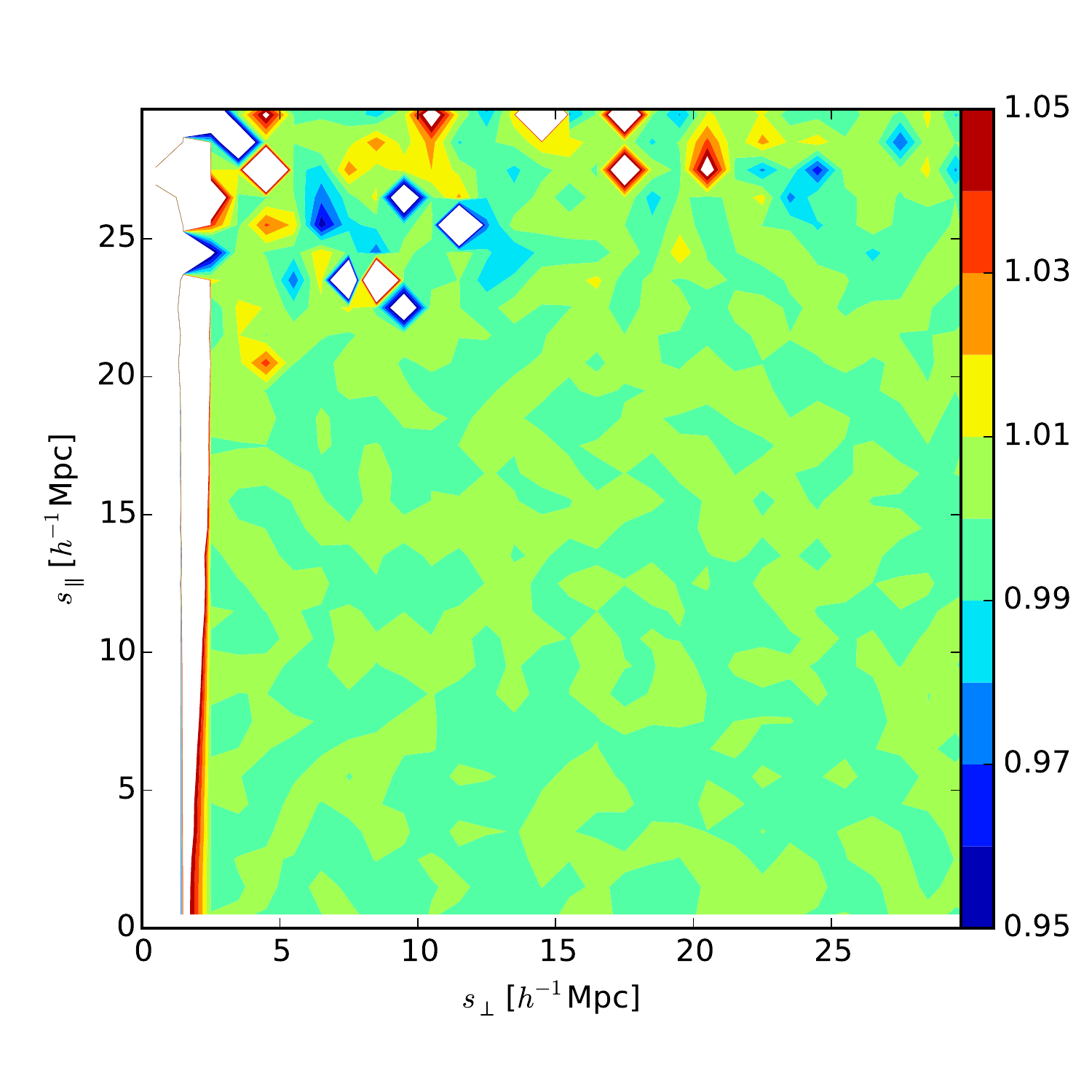}
   \includegraphics[width=8.5cm]{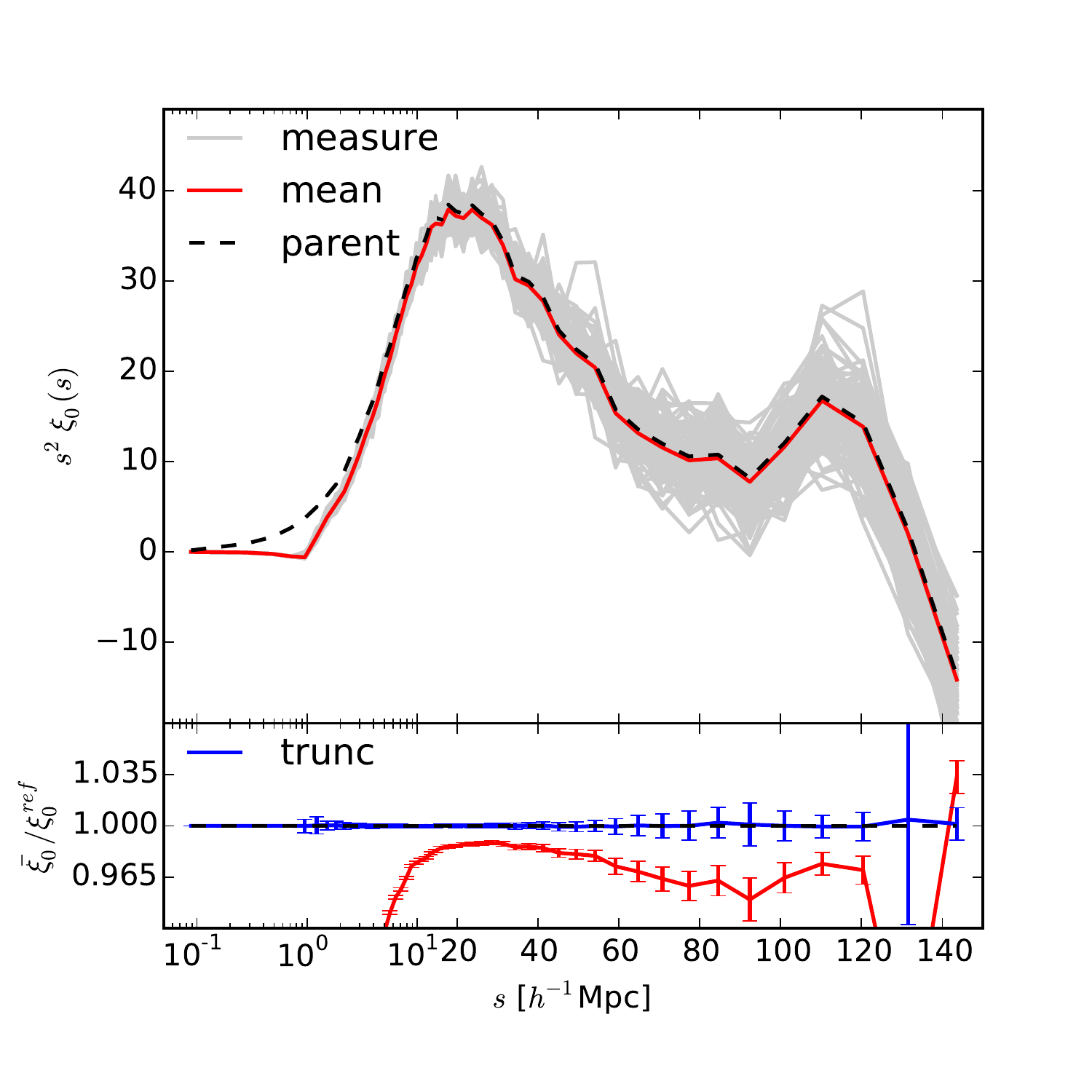}\\
   \vspace{-6mm}   
   \includegraphics[width=8.5cm]{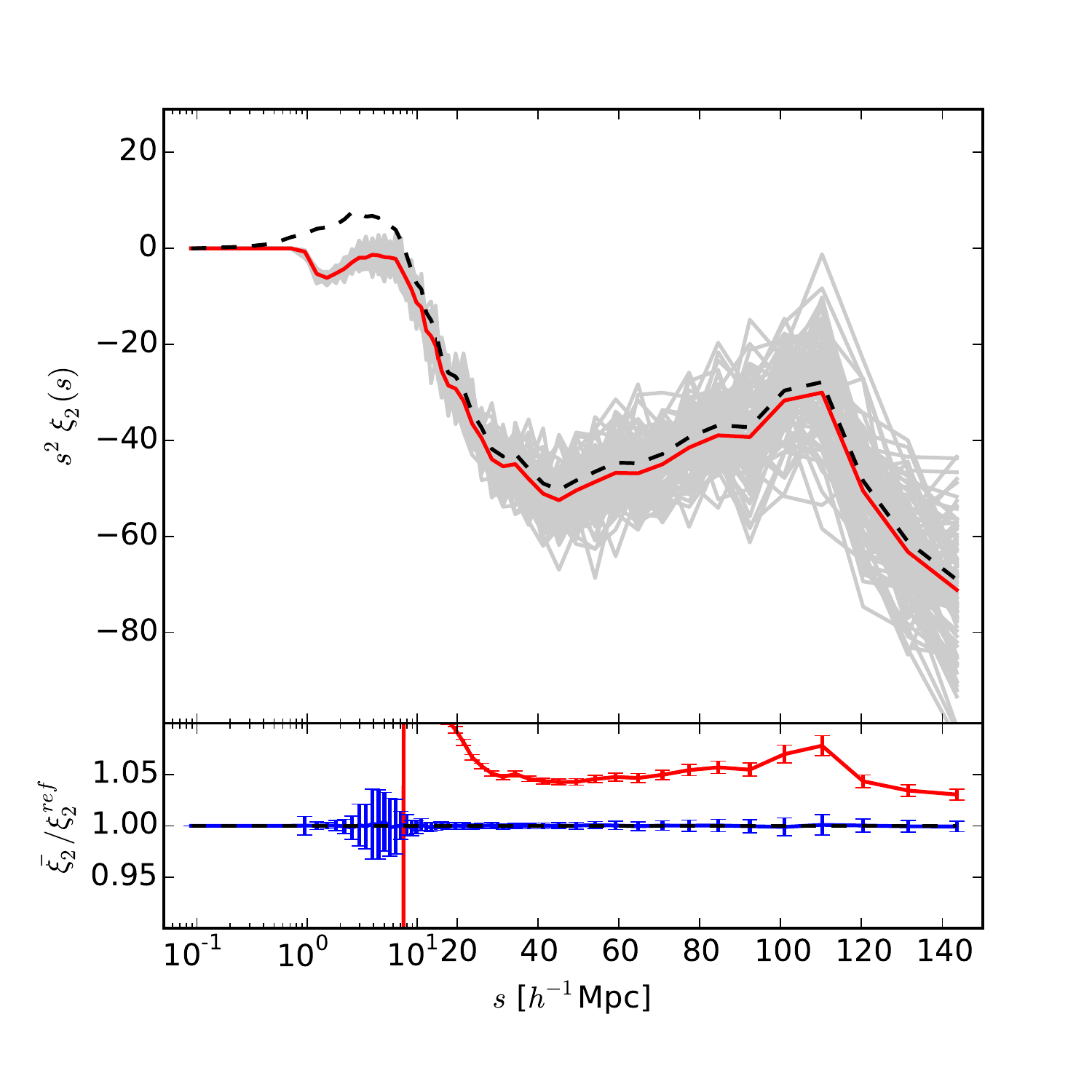}
   \includegraphics[width=8.5cm]{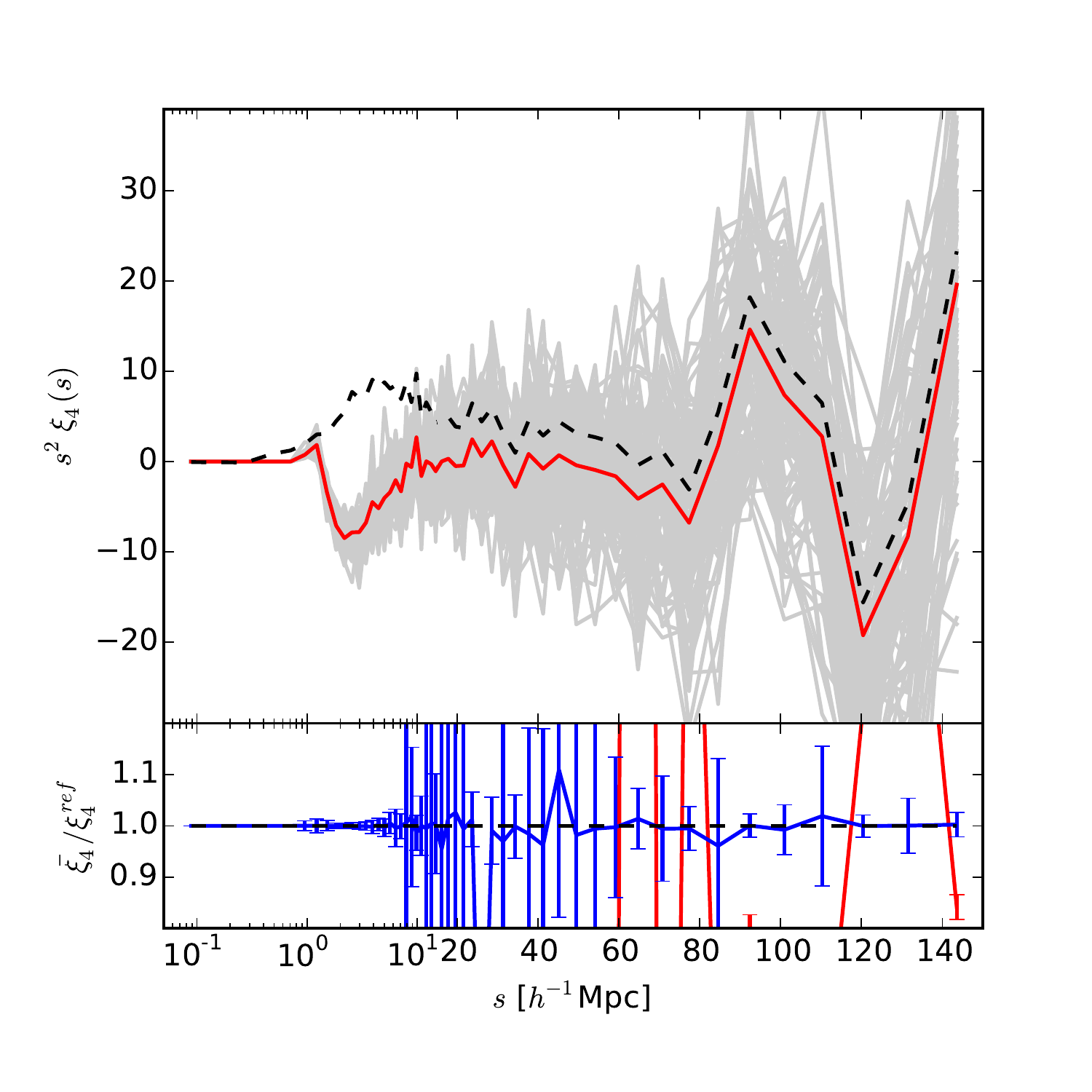}
   \caption{Performance of the PIP weighting scheme for the observing
     strategy OS1. Top left: ratio between the 2D correlation
     function $\xi(s_\perp, s_\parallel)$ obtained by averaging over
     992 realisations of the selection process and the reference value
     measured from the full parent sample. Top right: an assortment of
     measurements of the Legendre monopole $\xi_0(s)$ extracted from
     the 992 realisations, solid grey, and the total mean, solid red,
     compared to the reference value, black dashed. In the bottom
     frame we show the ratio between the mean and the reference value,
     solid red, together with the ratio between the mean and the value
     recovered from the parent sample when the
     $s_\perp < r_f = 1\mpcoh$ stripe is excluded, blue solid, with
     error bars of the mean. In order to show both the large- and
     small-scale clustering features at once, we adopt a logarithmic
     scale for the abscissa for $s_\perp < 15\mpcoh$ and linear
     elsewhere. Bottom left and bottom right: same as top right but
     for the quadrupole $\xi_2(s)$ and the hexadecapole $\xi_4(s)$,
     respectively.}
   \label{fig OS1}
 \end{center}
\end{figure*}
\begin{figure*}
 \begin{center}
   \includegraphics[width=8.5cm]{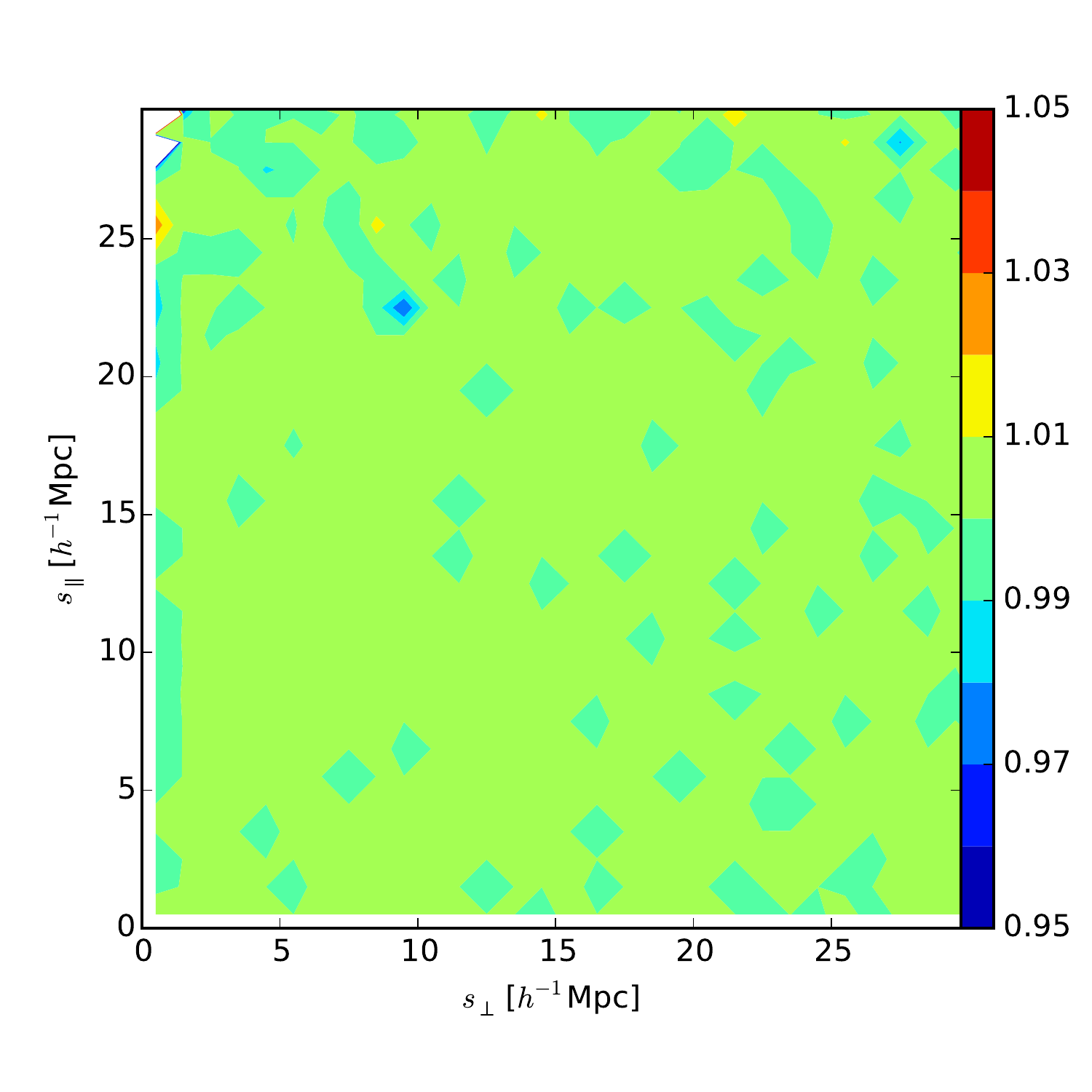}
   \includegraphics[width=8.5cm]{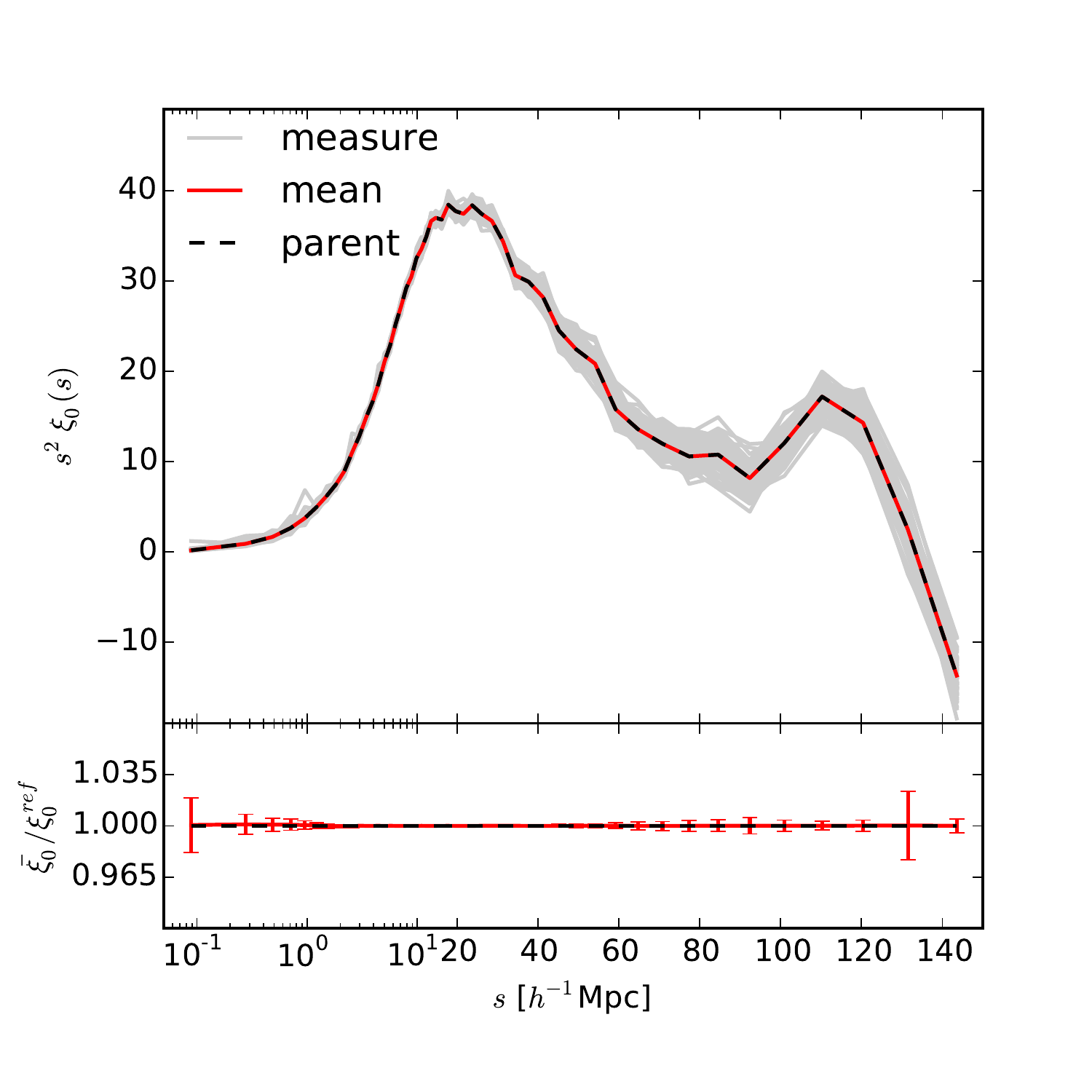}\\
   \vspace{-6mm}   
   \includegraphics[width=8.5cm]{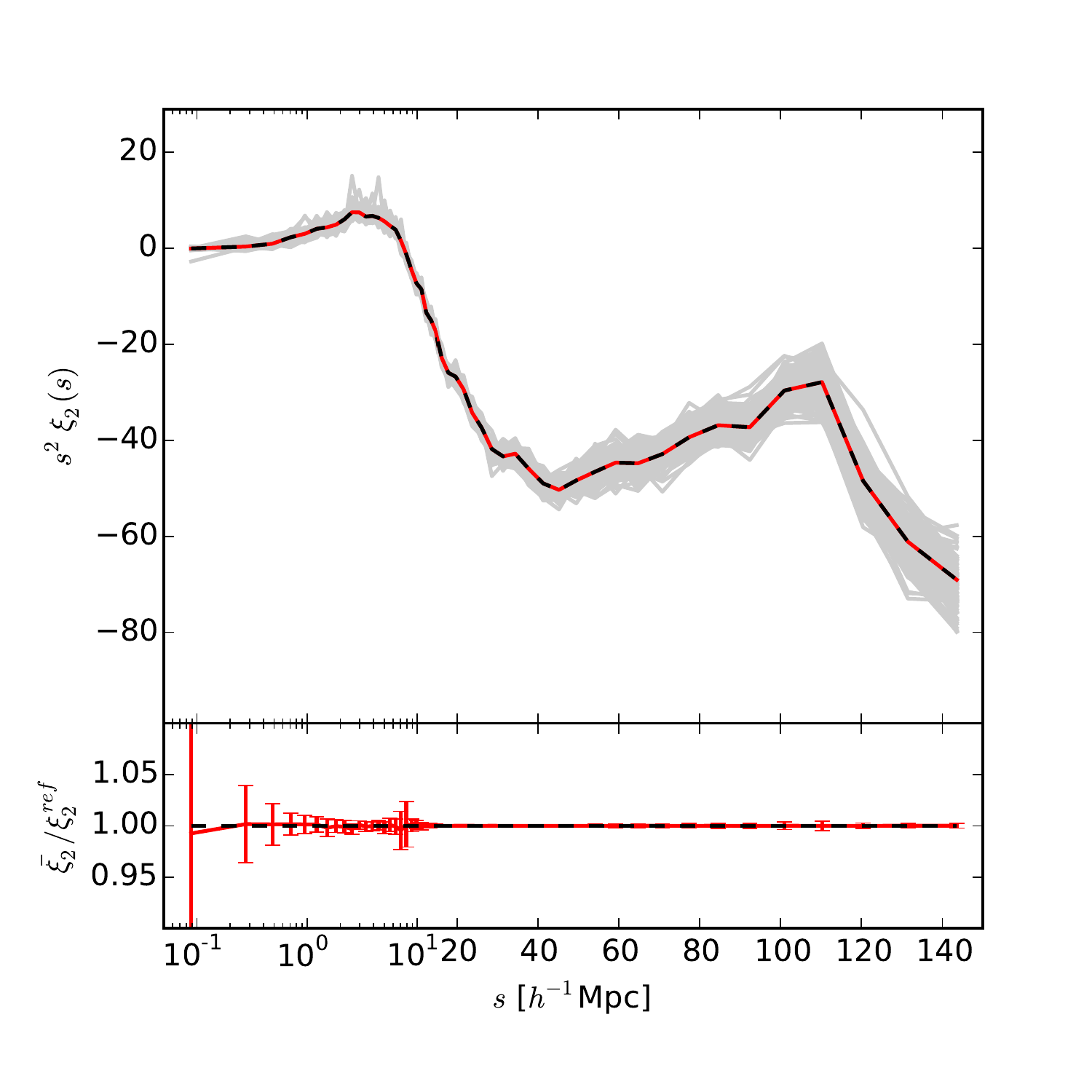}
   \includegraphics[width=8.5cm]{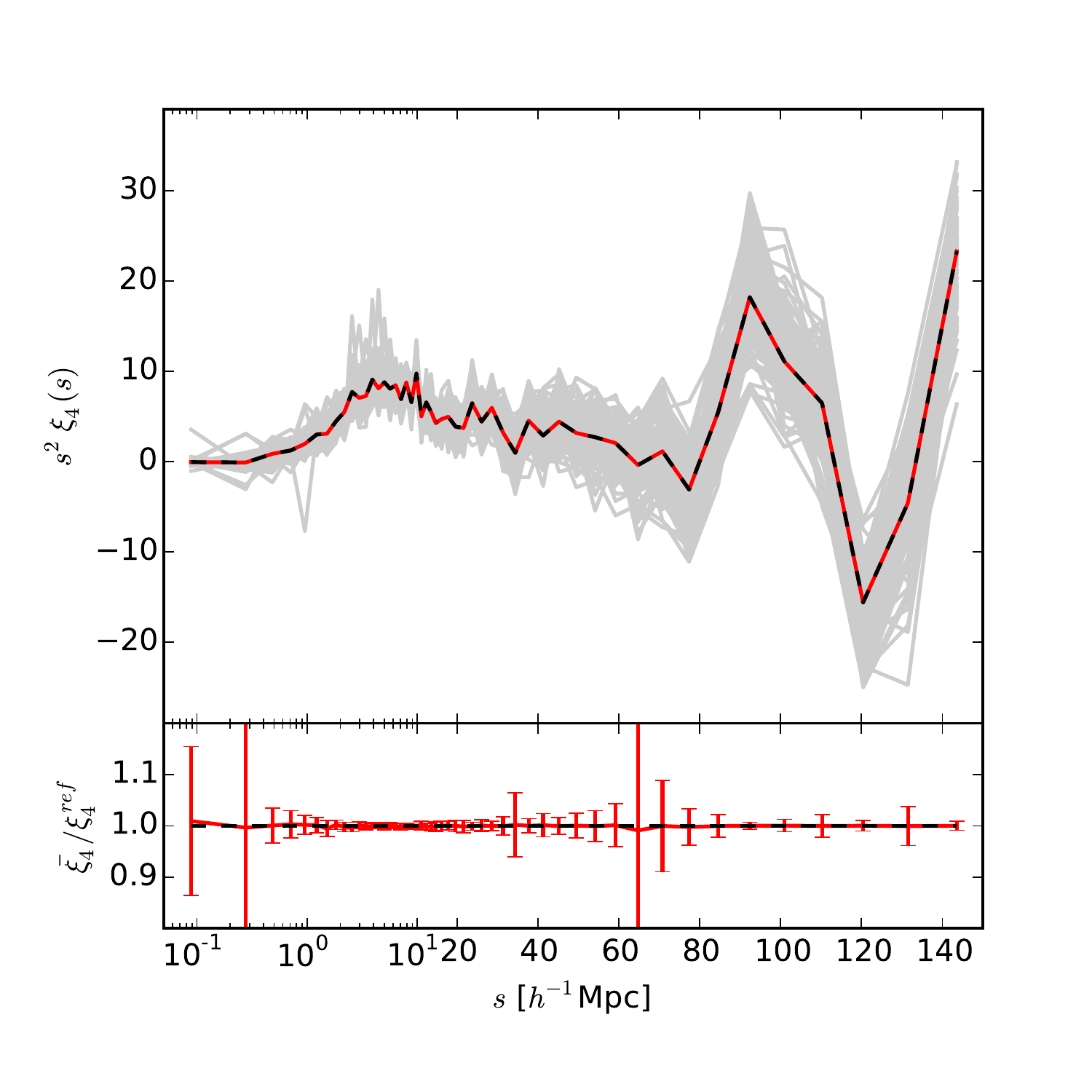}
   \caption{Same as Fig. \ref{fig OS1} but for observing strategy OS2.}
   \label{fig OS2}
 \end{center}
\end{figure*}
\begin{figure*}
 \begin{center}
   \includegraphics[width=8.5cm]{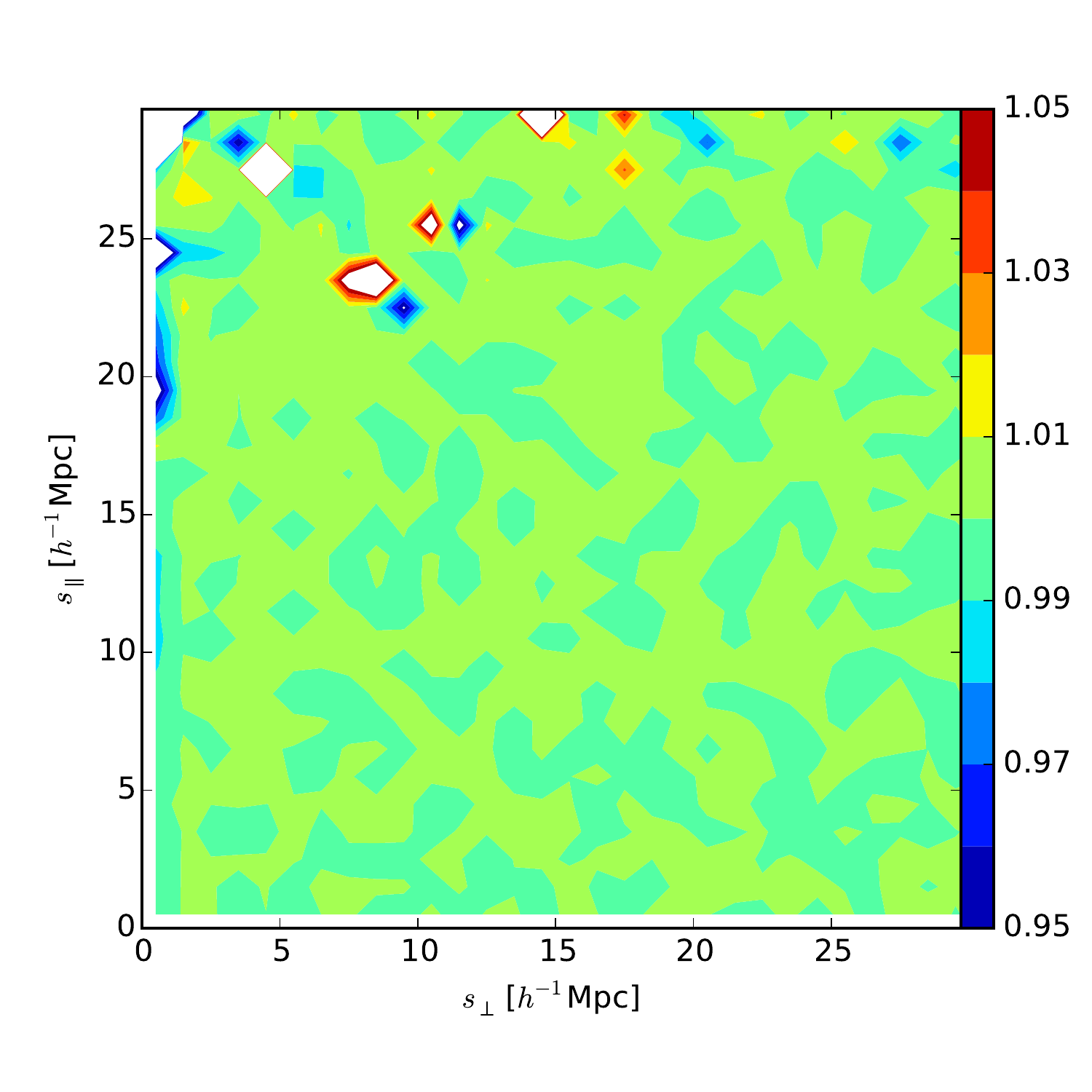} 
   \includegraphics[width=8.5cm]{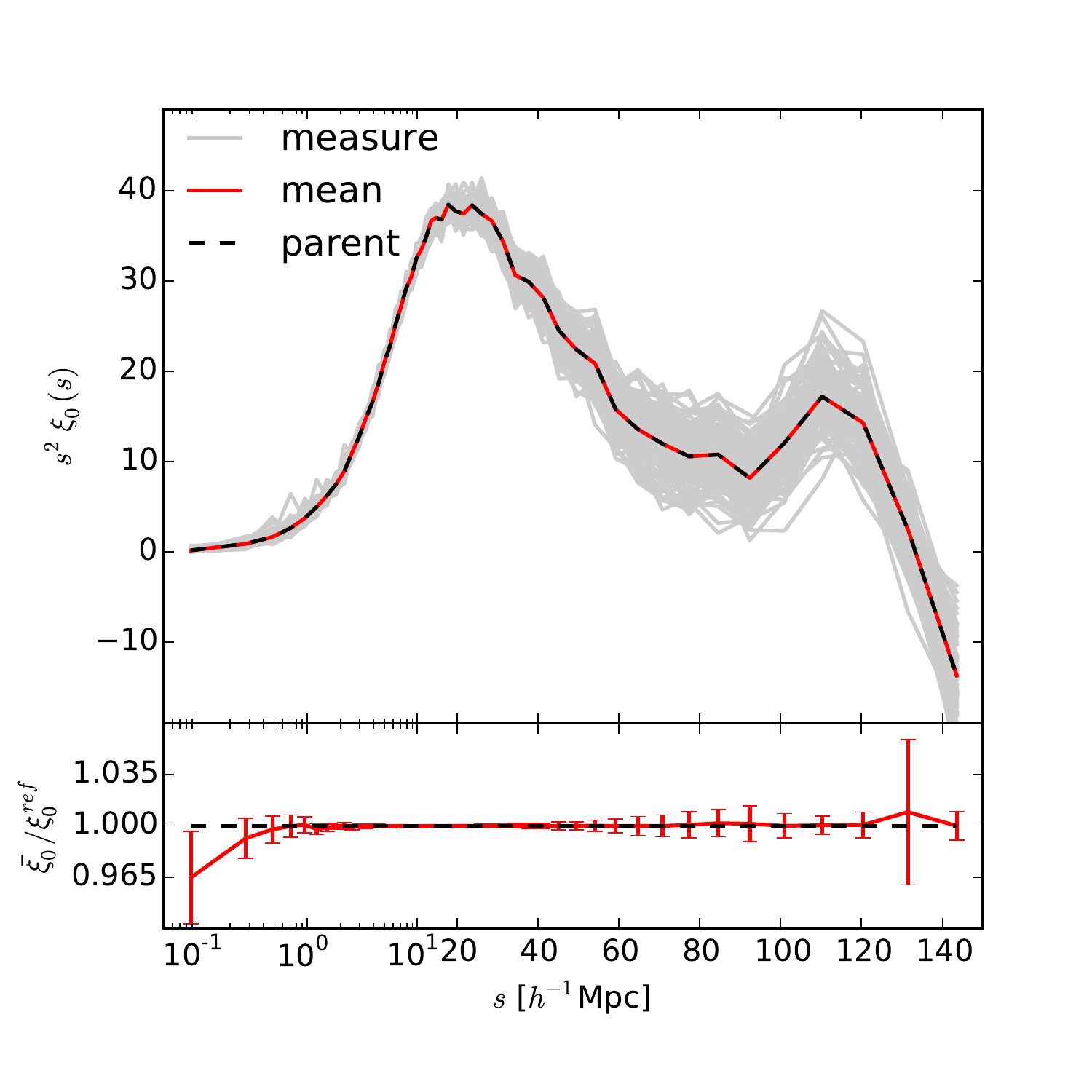}\\
   \vspace{-6mm}   
   \includegraphics[width=8.5cm]{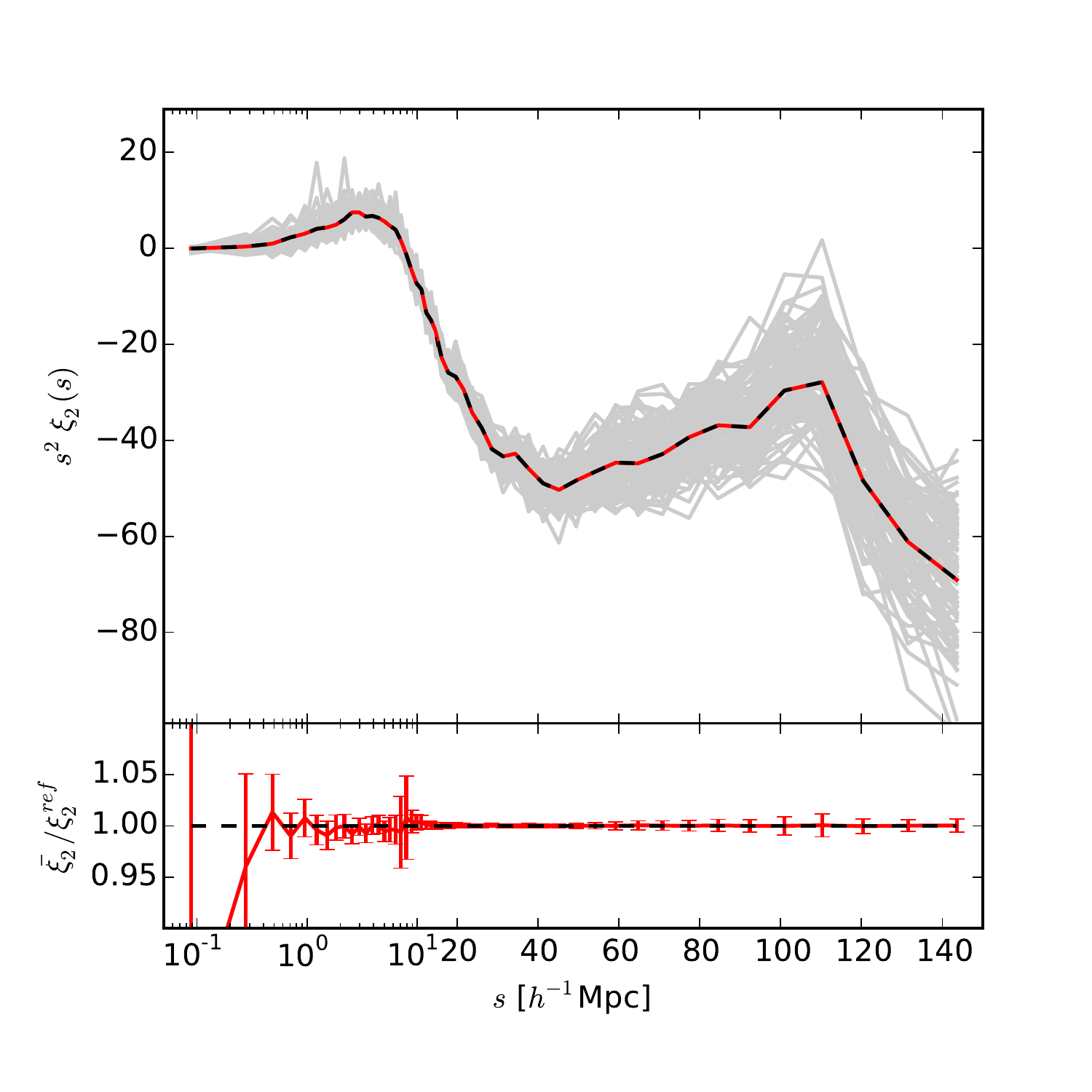}
   \includegraphics[width=8.5cm]{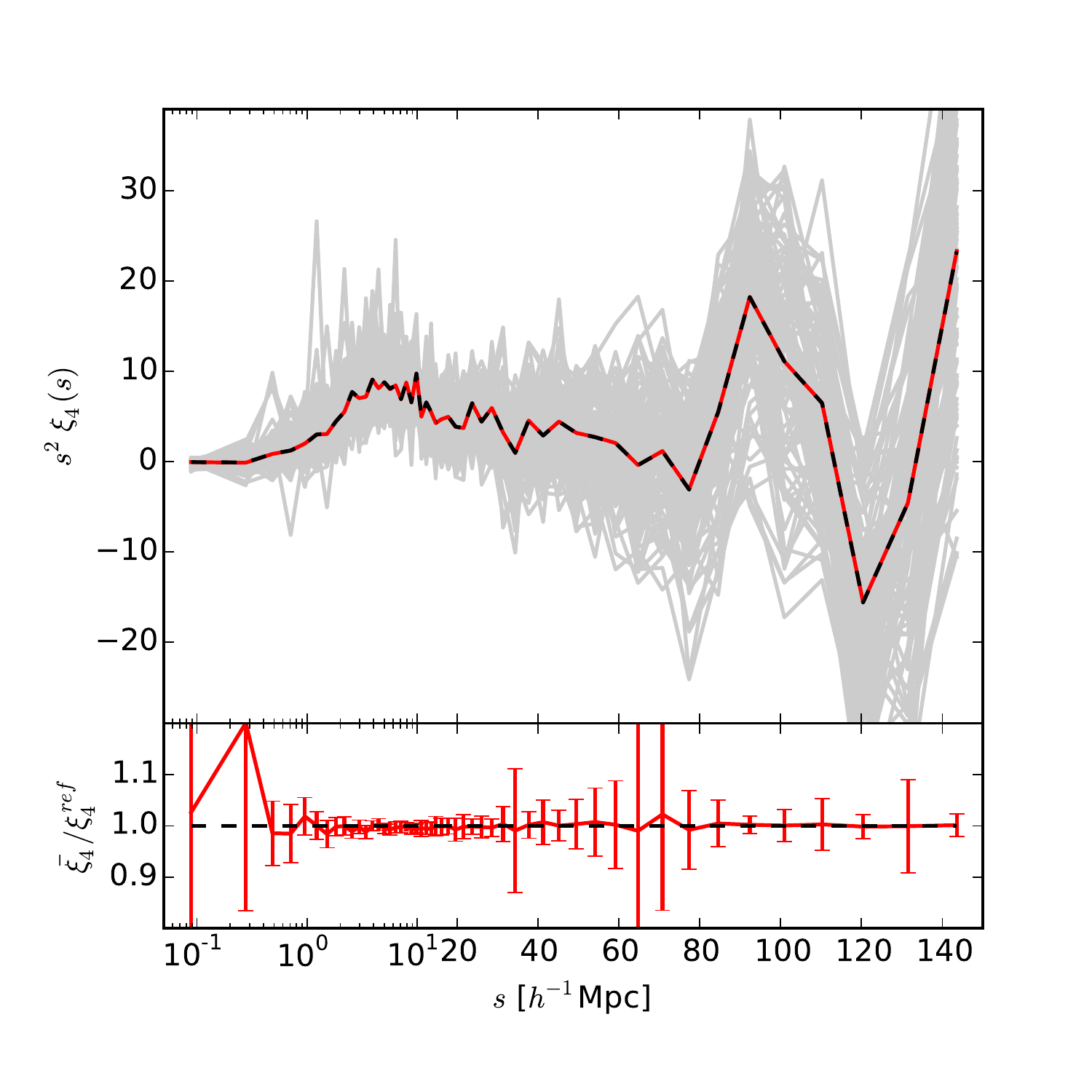}
   \caption{Same as Fig. \ref{fig OS1} but for observing strategy OS2sub.}
   \label{fig OS2sub}
 \end{center}
\end{figure*}
\begin{figure*}
 \begin{center}
   \includegraphics[width=8.5cm]{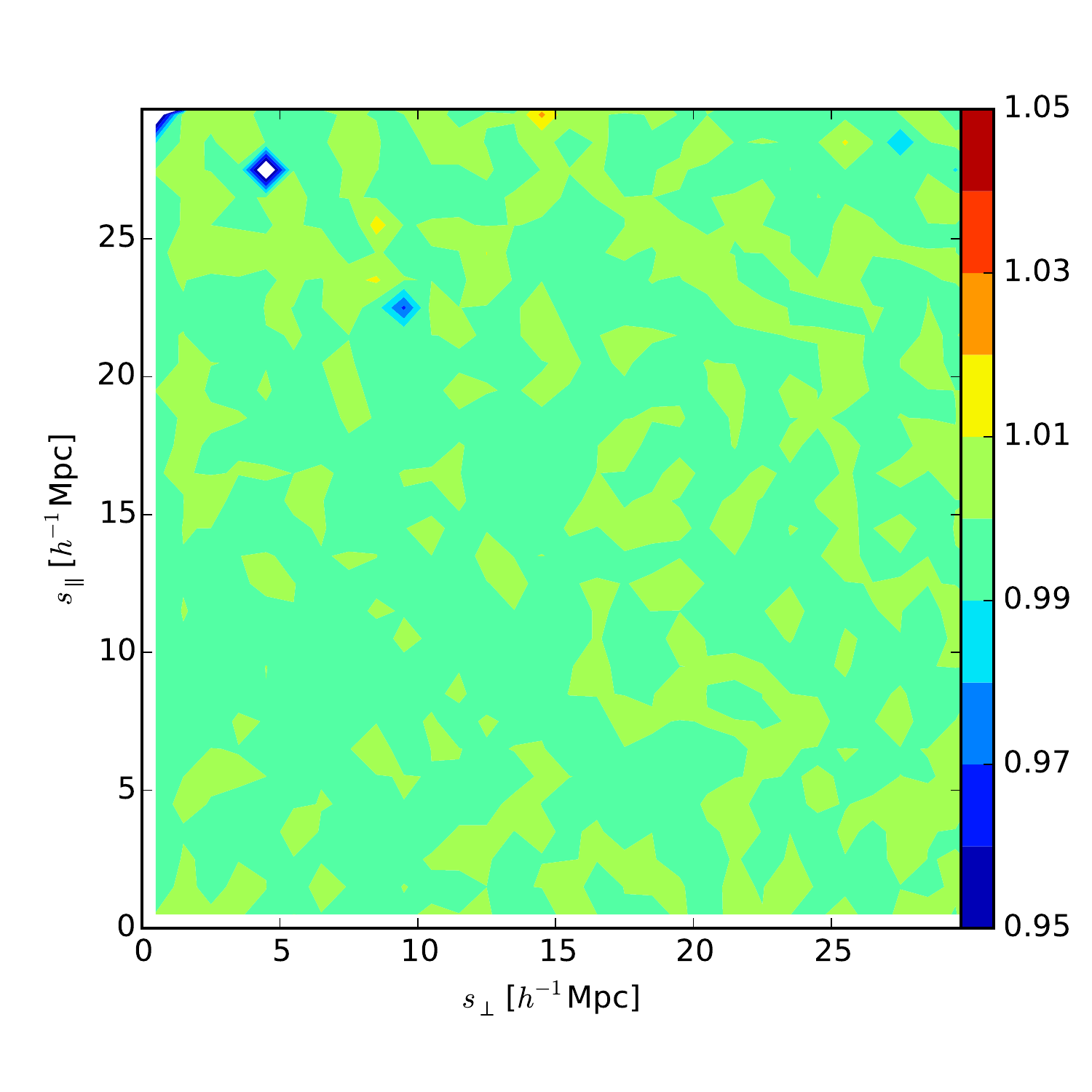} 
   \includegraphics[width=8.5cm]{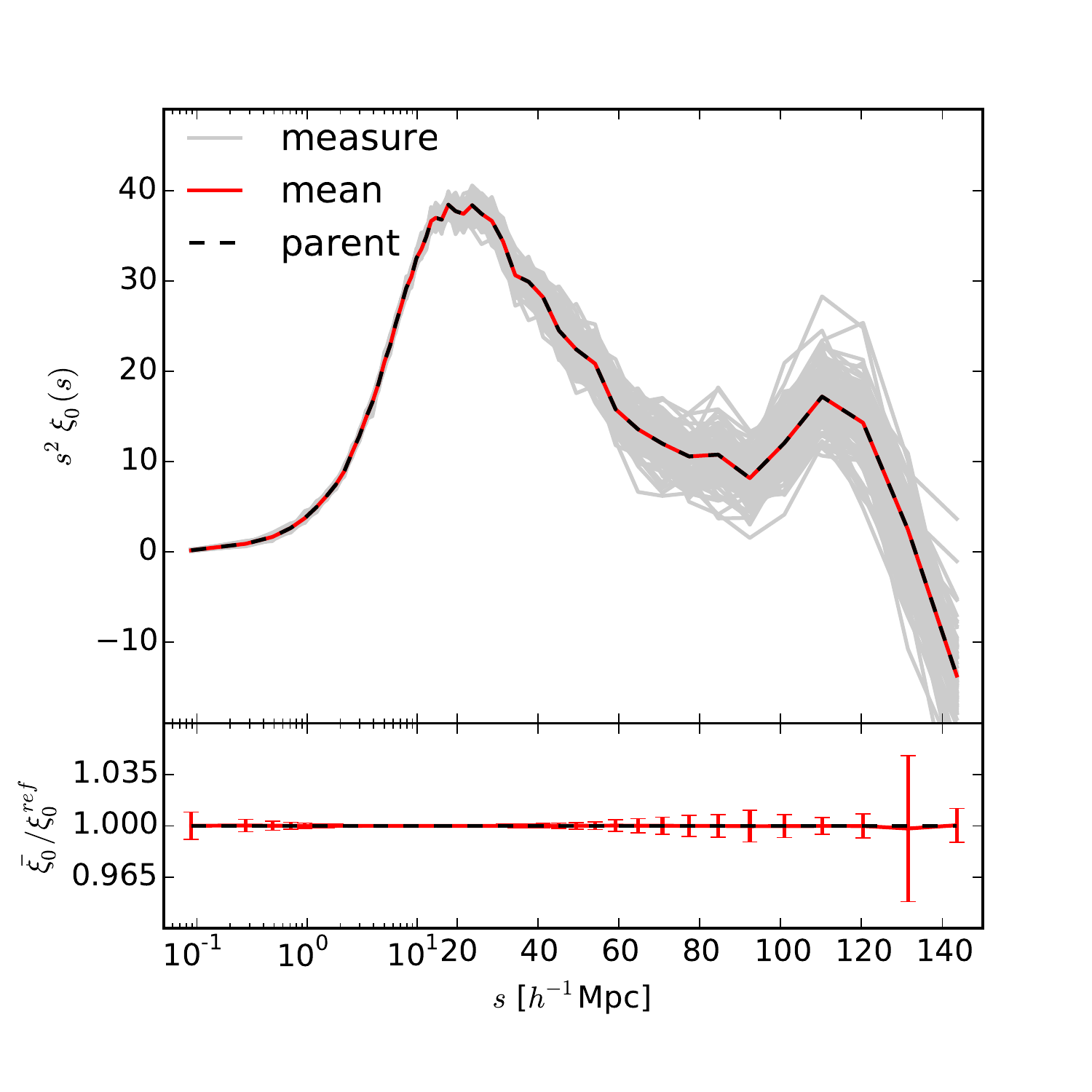}\\
   \vspace{-6mm}   
   \includegraphics[width=8.5cm]{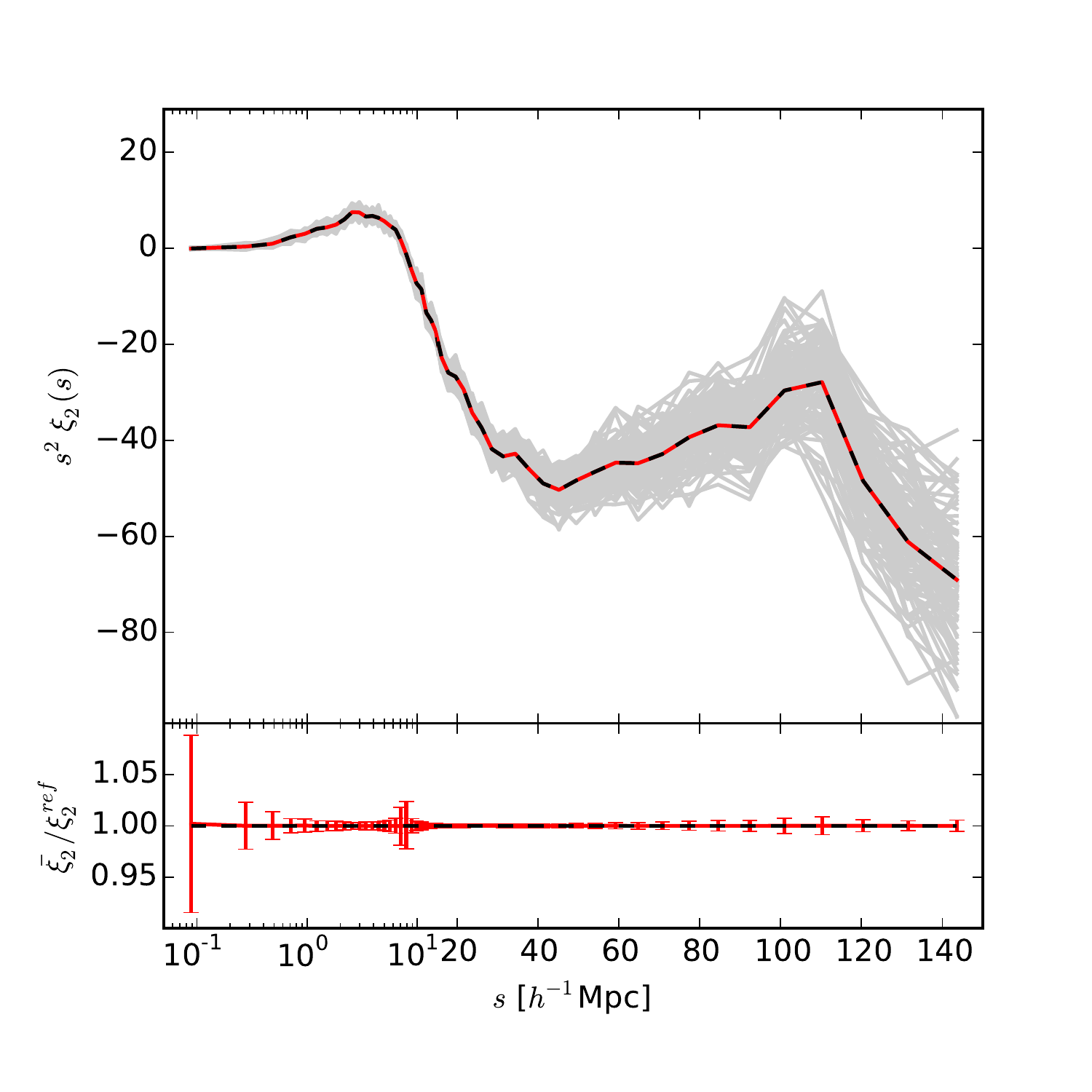}
   \includegraphics[width=8.5cm]{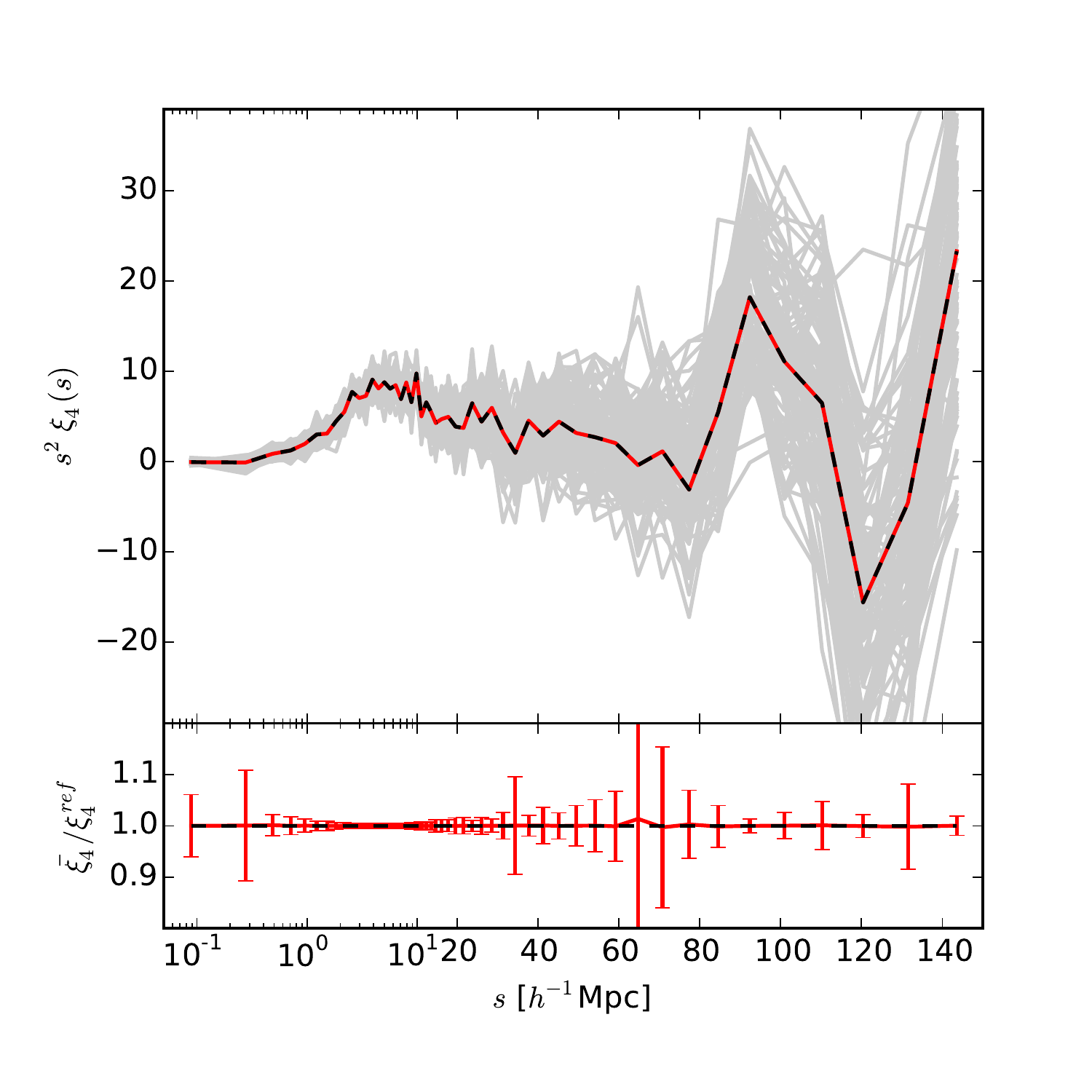}
   \caption{Same as Fig. \ref{fig OS1} but for observing strategy OSmulti.}
   \label{fig OSmulti}
 \end{center}
\end{figure*}
\begin{figure*}
 \begin{center}
   \includegraphics[width=8.5cm]{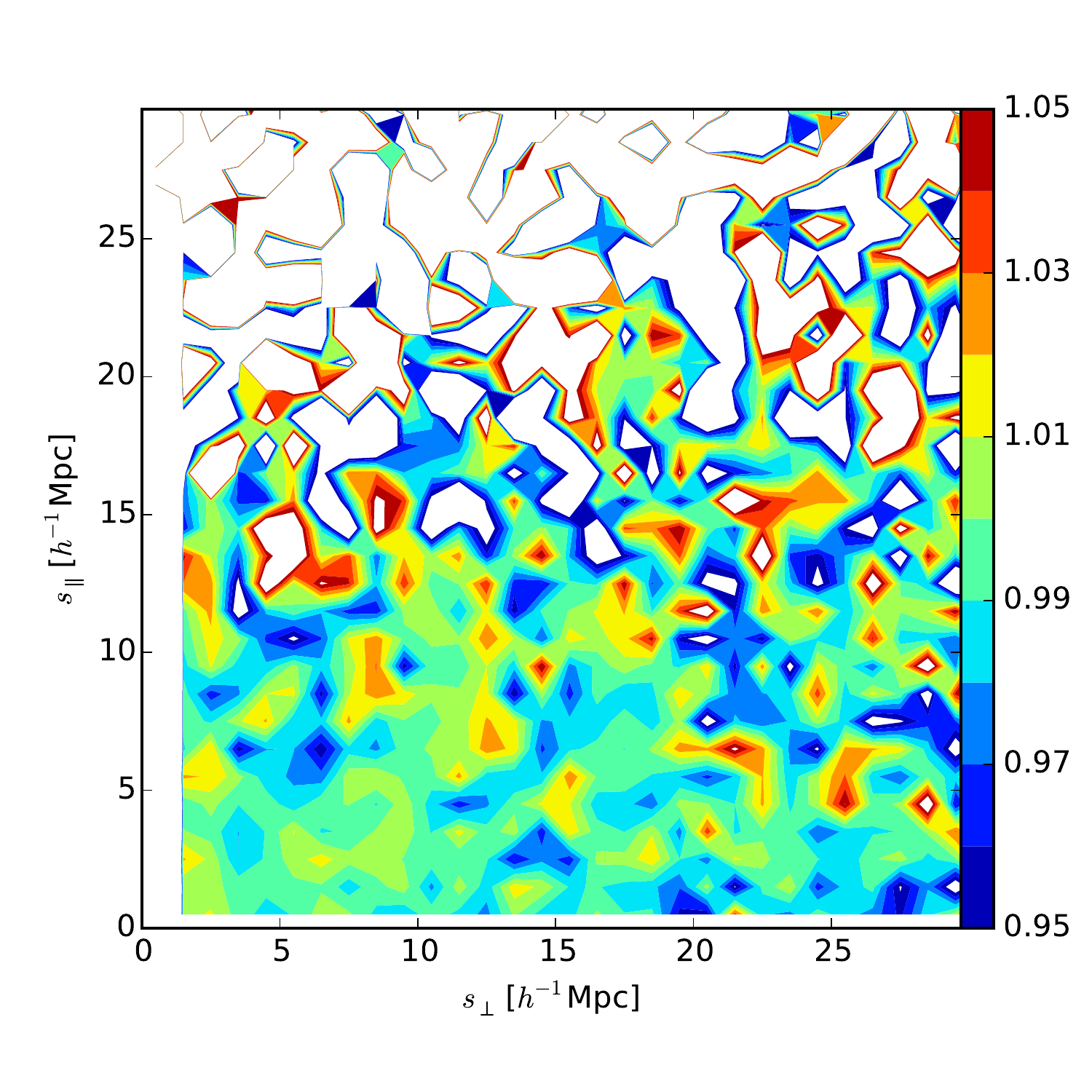} 
   \includegraphics[width=8.5cm]{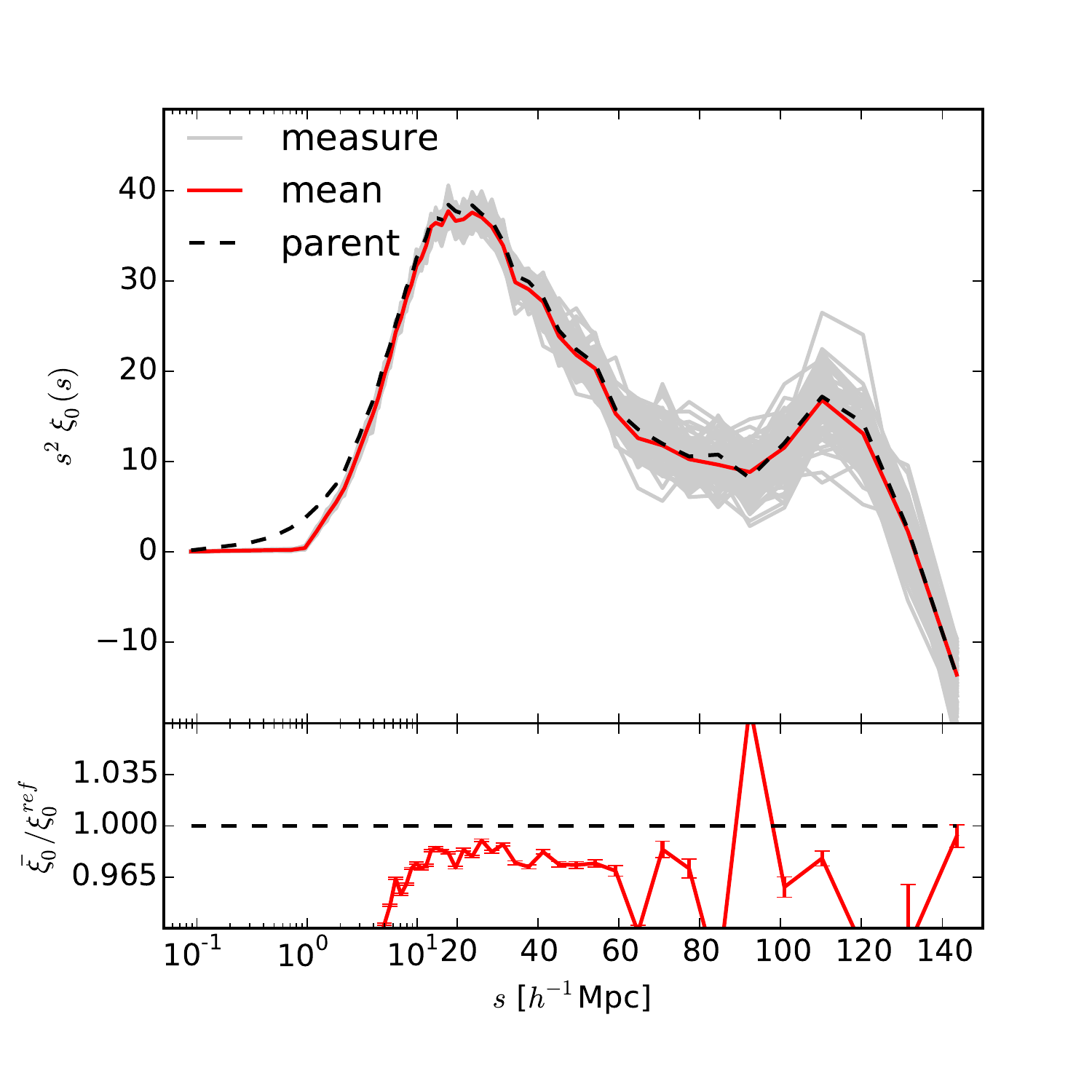}\\
   \vspace{-6mm}
   \includegraphics[width=8.5cm]{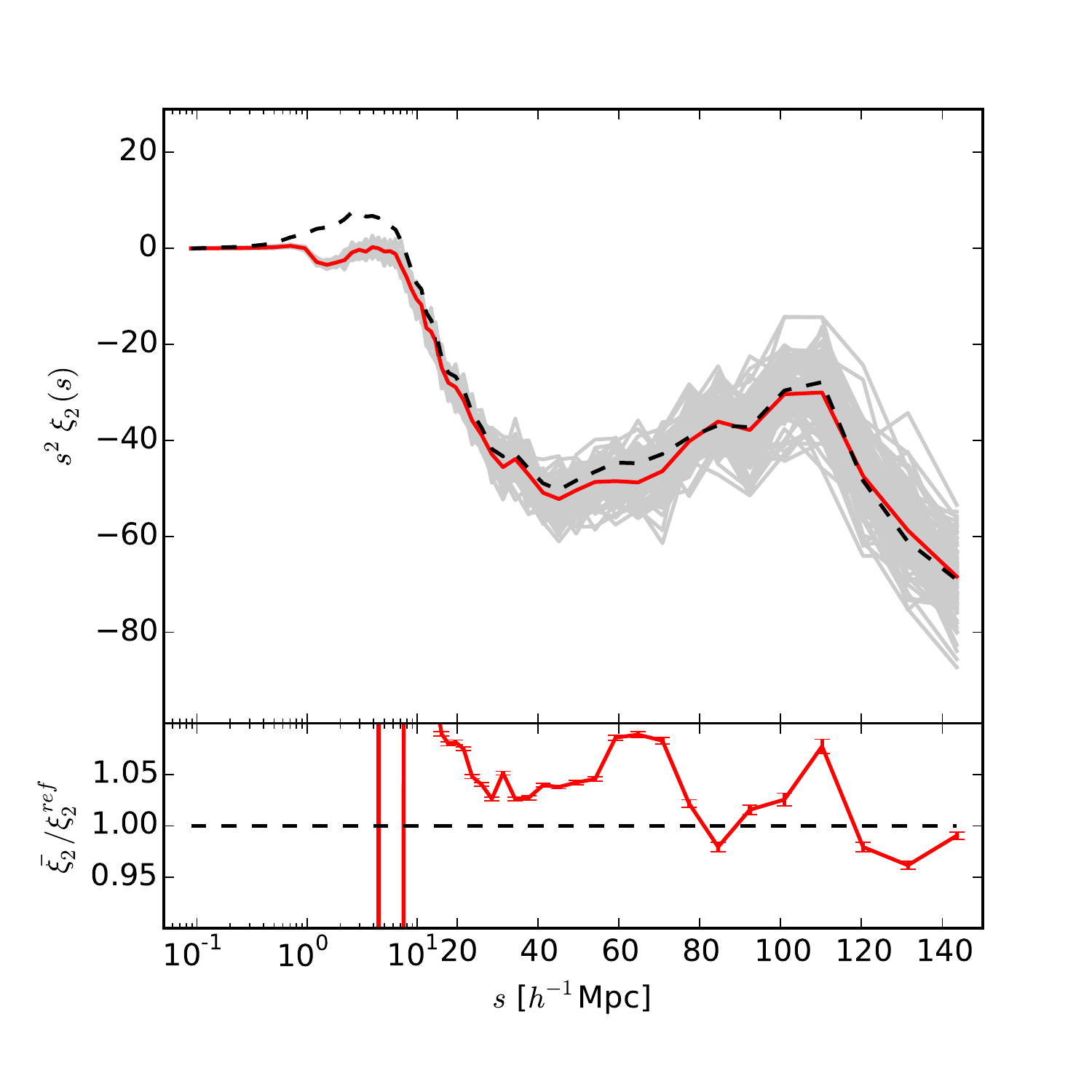}
   \includegraphics[width=8.5cm]{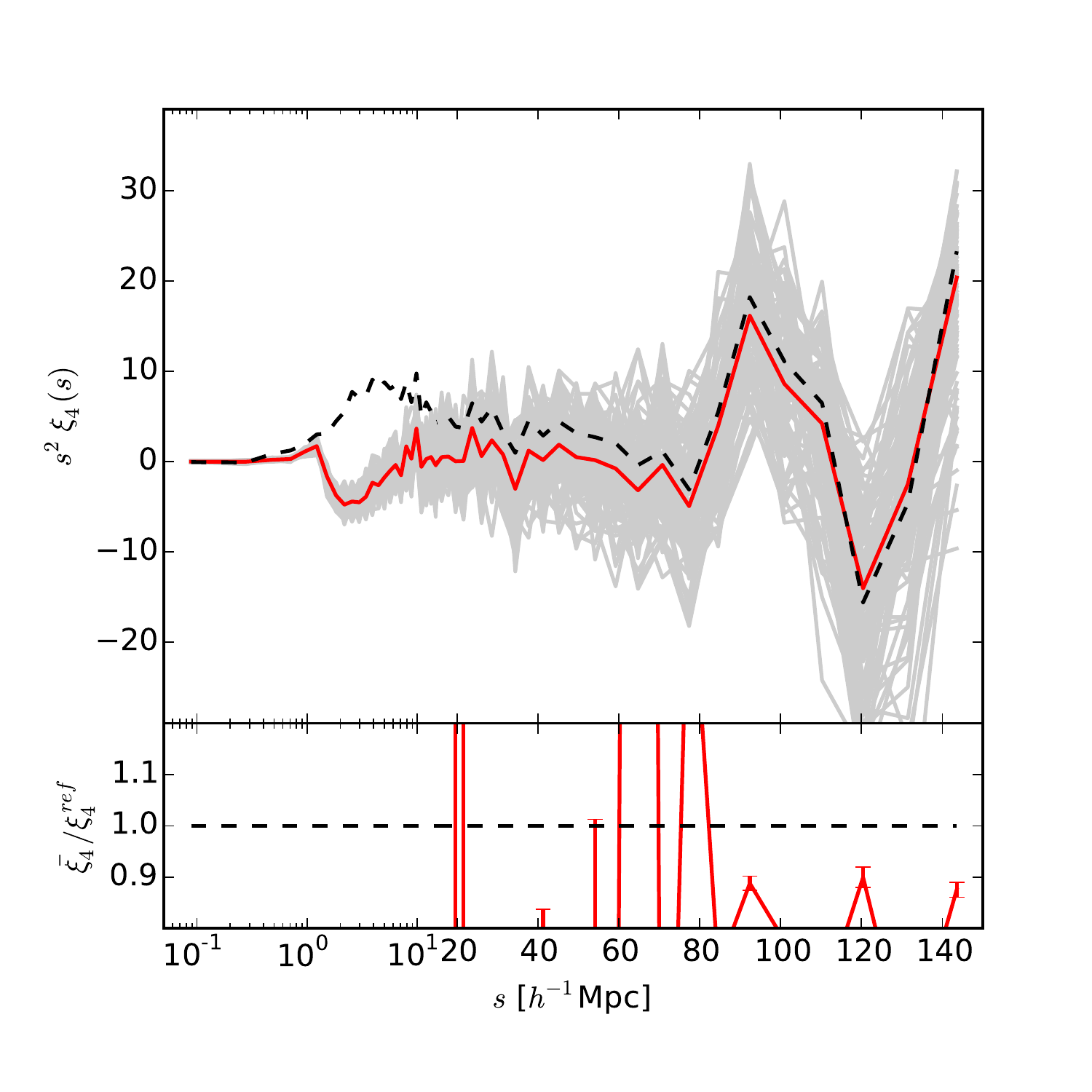}
   \caption{Same as Fig. \ref{fig OSmulti} but with NN assignment instead of PIP correction.}
   \label{fig NN}
 \end{center}
\end{figure*}
\begin{figure}
 \begin{center}
   \includegraphics[width=9.cm]{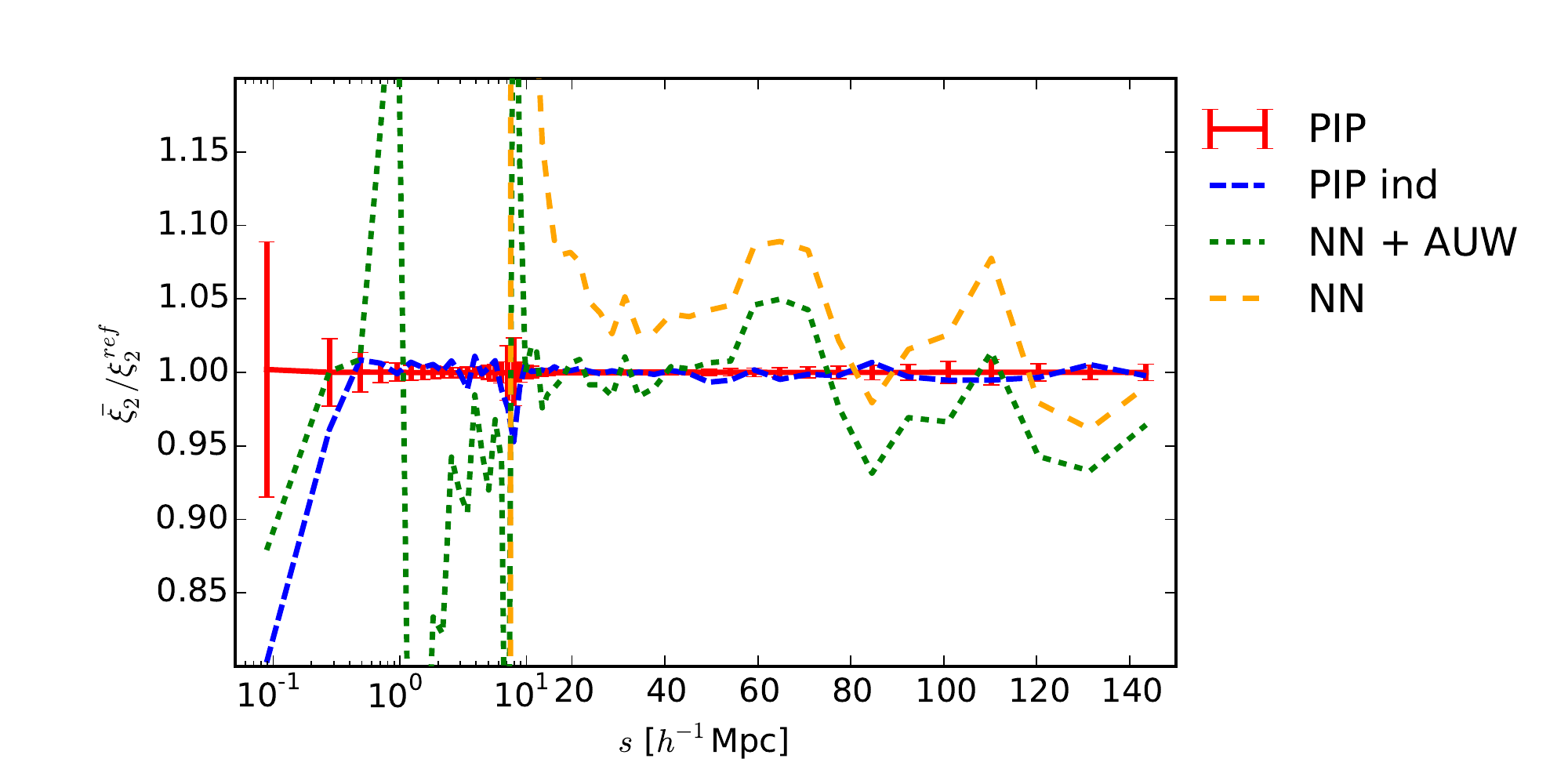} 
   \caption{Comparison of the accuracy of the quadrupole measured via
     PIP, PIP derived from and independent set of realisations of the
     selection process, NN, NN plus AUW, as labeled in the
     figure. Only OSmulti is considered.}
   \label{fig comp}
 \end{center}
\end{figure}
One of the properties that makes the PIP description preferable with
respect to more standard approaches is that it can be coupled to any
selection algorithm and parent sample, regardless, e.g., of the
selection correlation length.  Therefore, when testing the PIP weights
against simulations, we do not try to mimic any specific survey but
rather provide a general proof of concept for the method.  We use the
data from the MultiDark MDR1 run \citep{prada2012}, which adopts WMAP
cosmology,
$\{\Omega_m, \Omega_{\Lambda}, \Omega_b, \sigma_8, n_s\} = \{0.27,
0.73, 0.047, 0.82, 0.95\}$,
to describe the evolution of $2048^3$ particles over a
${(1000 \mpcoh)}^3$ cubical box.  For our analysis we apply a
$0.005\%$ dilution factor to the snapshot at redshift $z=0.5$.  The
resulting catalogue consists of $\sim 4.3 \times 10^5$ dark matter particles,
corresponding to a
$\sim 4.3 \times 10^{-4} \text{h}^3 \text{Mpc}^{-3}$ number density,
which is compatible with the actual number density of targets in a
modern galaxy survey.  In the following we sometimes refer to these
particles as galaxies and to this catalogue as the parent sample.  All
the collided catalogues we consider are obtained by applying, at least
once, the MR algorithm to the parent sample, in redshift space (we
assume plane-parallel approximation), with $r_f = 1\mpcoh$.
 
We first evaluate the effectiveness of our correction after a single
pass of the MR selection algorithm.  By repeating this observing
strategy, which we refer to as OS1, for different random seeds of the
algorithm, we have created 992 independent realisations all extracted from
the same parent sample (we discuss later the impact of keeping the
parent sample fixed).
  
The average fraction of galaxies that remain after this pass is $\bar{f}_s \sim 0.56$.
Only about half of them, i.e. one fourth of the total, can be classified as
uncollided because we have deliberately set an aggressive value of
$r_f$ in order to produce a significant effect. Each galaxy belonging
to this class has no other galaxies at separation smaller than $r_f$,
or, in other words, its probability of being selected after a single
pass is one.

In addition to $\bar{f}_s$, in Table~\ref{tab stats} we report the
fraction of discarded, collided and uncollided galaxies, for which we
adopt the subscripts $d$, $c$ and $u$, respectively.  Different
observing strategy are considered.  We also report $f_{px}$ and
$f_{p1}$, which represent the fraction of galaxies for which the
probability of being selected is $0<p<1$ and $p=1$, respectively.  By
construction none of the galaxies has selection probability $p=0$.
When OS1 is adopted, trivially $f_c=f_{px}$ and $f_u=f_{p1}$.

In the top left panel of Fig~\ref{fig OS1}, we show the ratio
$\bar{\xi}(s_\perp, s_\parallel)/\xi_p(s_\perp, s_\parallel)$ between
the average 2D correlation function measured via the PIP correction
from the 992 realisations and that of the parent sample.  PIP weights
are inferred from the same 992 realisations, as described in
Sec.~\ref{sec implement}, which means that we adopt
$N_{bits} = 992 = 31 \times 32$ bits.  This choice is arbitrary, based
on our checks, it seem likely that a significantly smaller number,
e.g. five times smaller, could be adopted for this quantity, if
needed.  However, the minimum acceptable $N_{bits}$ should be
determined according to the specifics of survey and selection
algorithm, which clearly goes beyond the purpose of this work.  For
the 2D correlation we focus on relatively small scales in order to
emphasise the $s_\perp < r_f$ stripe and because on larger scales, the
behaviour of $\xi$ becomes noisy\footnote{We could have shown the behaviour
  of the $DD$ counts, which is much more regular, it would not have
  been particularly informative though, since on large scales pair
  counts are completely dominated by the geometry and a $1\%$
  systematic error on $DD$ might translate into a $100 \%$ one on
  $\xi$.} due to the small bin sizes.  See App.~\ref{sec details} for
details on how we measure the different statistics and the
corresponding binning.  As expected, the PIP correction provides an
unbiased estimate on all scales $s_\perp > r_f$.  In the top right
panel we explicitly show the Legendre monopole $\xi_0(s)$ measured
from the various realisations, grey solid, together with the mean
$\bar{\xi}_0$, red solid, and that measured from the parent sample
$\xi_0^{ref}$, black dashed.  At the bottom of the same panel we
report the ratio $\bar{\xi}_0/\xi_0^{ref}$ with corresponding error
bars of the mean.  Similarly, in the bottom panels we show the
behaviour of the Legendre quadrupole $\xi_2(s)$ and hexadecapole
$\xi_4(s)$.  In order to properly visualise the impact of fibre
collisions on all the scale of interest, for these plots we adopt a
logarithmic scale for $s<15\mpcoh$, which becomes linear at larger
separations.  All of the multipoles are clearly affected by systematic
bias, which grows with the order of the multipole.  This is not
surprising at all since the evaluation of the multipoles requires
$\xi(s,\mu)$ to be integrated over the range $0 \le \mu \le 1$, where
$s=\sqrt{s_\perp^2+s_\parallel^2}$ and $\mu=s_\parallel / s$.  As a
consequence the $s_\perp < r_f$ stripe, in which the measurements are
unavoidably uninformative when OS1 is adopted, affects the accuracy of
our estimates on all scales.  In practice, this problem can be easily
circumvented just by excluding such band from the integration,
i.e. using truncated multipoles \citep[see e.g.][]{reid2014,
  mohammad2016}.  In order to prove this, in addition to the ratio
$\bar{\xi}_n/\xi_n^{ref}$, we also report $\bar{\xi}_n/\xi_n^{trunc}$,
blue solid, where $\xi_n^{trunc}$ are the multipoles recovered from
the parent sample using only the relevant scales, $s_\perp > r_f$.
Clearly the systematic effect is removed.

Next, we consider the possibility of running a second pass of
observations, in which the MR algorithm is applied to the galaxies
discarded after the first pass (observing strategy OS2).  With this
new strategy we obviously see a large improvement in the estimate of
the 2pt statistics, Fig.~\ref{fig OS2}.  Specifically, after PIP
correction, $\xi(s_\perp,s_\parallel)$ is now unbiased on all
scales (top left panel), as expected, i.e. we now have pairs and can estimate $\xi$ for all $s_\perp$ and $s_\parallel$.
Consequently, all the multipoles are unbiased on all scales, as well.  Also, the variance is
significantly reduced with respect to OS1.  It is interesting to note
that when a complete second pass is performed, the fraction of
galaxies that are always selected grows from $f_{p1}=0.25$ to
$f_{p1}=0.51$, see Tab.~\ref{tab stats}.  This tells us that most of
the collided galaxies belongs to AFOF structures more complex than
simple pairs, otherwise we would have $f_{p1}\sim1$ and PIP weighting
would be essentially equivalent to the NN correction (at least for the
MR algorithm).

We now explore a scenario in which only a subset of the full survey
area is observed twice.  First, we consider the case of this subset
being a square of $500\mpcoh$ side, i.e. $1/4$ of the total area,
observing strategy O2sub.  For each of the 992 realisations we centre
the square randomly and run a second pass of the MR algorithm on the
galaxies inside it.  As discussed in Sec.~\ref{sec small scales}, it
is convenient to model the positioning of the second-pass area as a
stochastic process\footnote{The distribution does not necessarily have
  to be uniform.} because, by doing this, we enforce each pair to have
non zero probability of being observed, which is a crucial property in
building all-scale unbiased estimators.  From Fig.~\ref{fig OS2sub} we
can see that all our measures remain unbiased, as for OS2.  With
respect to this latter the variance is increased, which is a trivial
consequence of having fewer pairs and an increased shot noise.

We now consider the possibility of a more complex geometry for the
second-pass area.  Specifically, we split the $500\mpcoh$ square
into 100 smaller squares of $50\mpcoh$ side (observing strategy
OSmulti).  Since we implement this strategy just by setting a smaller
square size and iterating 100 times OS2sub, some of the squares
overlap.  As a consequence, the total second-pass area is smaller on
average.  On the other hand, overlap regions are observed more than
twice, meaning that the 3D clustering inside them is almost
perfectly known.  We see from Fig. \ref{fig OSmulti} that OSmulti
yields overall similar results to OS2sub, in terms of both precision
and accuracy.  More in detail, we note an improvement on scales
$s \lesssim 10\mpcoh$, which we can attribute to the additional
information on the small-scale clustering coming from overlap regions.
This improvement seems not to come at the cost of any degradation of
the large-scale signal.

For comparison, in Fig.~\ref{fig NN} we show what happens if instead
of the PIP weights we adopt the standard NN correction.  We only
report results for OSmulti, but the behaviour does not significantly
depend on the observing strategy.  The estimate of the clustering
obtained via NN assignment is clearly less accurate than that obtained
via PIP.  Focusing on the behaviour of the multipoles, we note some
similarities with the OS1 scenario previously discussed (Fig.~\ref{fig
  OS1}).  This suggests that part of the observed bias comes from the
lack of small-separation pairs, which is an unavoidable problem when
using a pure NN scheme.  It is nonetheless clear from the large scale
oscillations in the ratio $\bar{\xi_n} / \xi_n^{ref}$ that finding a
correction for this effect would not be enough to match the
performance of PIP weighting.  This is not surprising since, as
discussed in Sec. \ref{sec range}, NN assignment can be seen as an
approximate way to evaluate the selection probability of the pairs.

For all the PIP measurement reported in this section we applied AUW,
as discussed in Sec.~\ref{sec var}.  As expected, this helped in
reducing the variance of our estimator on large scales, where a $1\%$
fluctuation in $DD$ can be easily amplified to a $100\%$ fluctuation
in correlation function.  The improvement is relevant for the monopole
in the OS2sub and OSmulti case, i.e. when clustering information is
extrapolated from a subset to the total area.  Part of this is due to
the fact that, while 992 bits (i.e. realisations) are sufficient to
sample the local effect of the AFOF structures on the selection
probability, they are not enough to accurately sample collective
effects coming from the positioning of the second-pass area, whose
distribution is know to be uniform by construction.  Furthermore, even
with a larger number of bits, the intrinsic coupling between PIP
weighting and clustering would tend to emphasise the normal
fluctuations of this latter from one second-pass area to another.
Luckily, both this effects are efficiently counterbalanced by the AUW
correction.
 
Similarly to what we did for PIP weighting, we have considered the
possibility of applying simultaneously AUW and NN assignment.  In this
case AUW mostly acts to correct some of the small-scale issue
discussed above for the NN procedure.  Clustering estimates are
improved with respect to Fig. \ref{fig NN} but not enough to be
unbiased at a level of precent precision on all scales, as expected.
For an example of the effect of this correction on the behaviour of
the quadrupole in the OSmulti case see Fig.~\ref{fig comp}.

We are using the same targeting realisations to calculate the PIP weights as we are using to measure mean and variance of the multipoles.
Because of direct cancellations in the pair counts, the scatter seen is reduced from the scatter if these were independent realisations.
This can be seen in Figs.~\ref{fig OS1}-\ref{fig OSmulti}, where the data are not fluctuating within the error bars.
We have tested this by creating a new set of 992 targeting realisations,
obtained by running the MR algorithm on the same parent sample but with different random selection of observations.
From this new set we obtained an estimate of the PIP weights, which we then used to measure the clustering with the independent set of realisations (i.e. the same same set we used for Figs.~\ref{fig OS1}-\ref{fig OSmulti}).
As expected, we obtained very similar results to those reported in Figs.~\ref{fig OS1}-\ref{fig OSmulti}, but with a scatter in the ratios $\bar{\xi}_n / \xi_n^{ref}$ that is more compatible with the error bars reported (see Fig.~\ref{fig comp} as an illustrative example).

The decision to keep the parent sample fixed is obviously meant to
match a real scenario in which, given a catalogue with only angular
positions, we have to choose which galaxies to observe
spectroscopically.  In this scenario, the PIP weights to be applied to
the data can be obtained following the same procedure we have adopted
here.  However, it is important to discuss what happens when we
consider the stochasticity of the parent sample, especially when
evaluating covariance matrices for the clustering statistics.
Depending on the selection algorithm and on the characteristics of a
survey, volume and number density in particular, the contribution to
the variance coming from a fibre-collision-like problem with respect
to the intrinsic cosmic variance can be negligible, comparable or
dominant (but we can always think of a scale below which it becomes
dominant).  The error bars reported in this section, refer to the
latter case, e.g. when the survey's volume is very large.  In the case
in which the problem is negligible we just resort to a set of mock
catalogues and the expected value for $DD$ is trivially given by
$N_{mocks}^{-1} \sum_{m=1}^{N_{mocks}} DD_m$, where $N_{mocks}$ is the
total number of mocks.  If instead, the two contributions are
comparable we have to run the selection algorithm $N_{bits}$ times on
each mock to derive the corresponding PIP correction.  The expectation
value becomes
$N_{mocks}^{-1} \ N_{bits}^{-1} \ \sum_{m=1}^{N_{mocks}}
\sum_{n=1}^{N_{bits}} DD_{mn}$,
where $DD_{mn}$ are computed via PIP weights\footnote{Note that the
  number of samples that we actually need to store and for which we
  have to perform pair counts, does not necessarily have to be
  $N_{mocks} \times N_{bits}$.  In order to save computational
  resources, for the evaluation of covariance matrices we are free to
  use $N_{mocks} \times N_{eff}$, with $N_{eff} < N_{bits}$, since
  $N_{bits}$ is just a choice for the precision of the PIP sampling.}.
By the same reasoning used for the single-parent-sample case, we
see that $\langle DD \rangle = \langle DD_p \rangle$, i.e. PIP
correction yields unbiased clustering estimates even when the
stochasticity of the parent sample is taken into account, but,
obviously, the covariance changes.

\section{Conclusions}\label{sec conclusions}

We have developed an unbiased estimator for the galaxy clustering in
the presence of correlated missing observations.  The method relies on
the concept that the stochastic process with which the catalogue of
observed galaxies is extracted from a parent sample is known and can
be simulated. This is clearly the case when dealing with the fibre
collision issue, for example.  We have shown that by weighting each pair by
its pairwise inverse probability (PIP) of being observed, the correct
two-point correlation function is recovered on all scales for which there
are no pairs with null selection probability.  For observing
strategies with overlap regions, this translates into all-scale
unbiased measurements, whereas for one-pass surveys, the
zero-probability region is basically uninformative and can be easily
excluded from the analysis without information loss.

By introducing the concept of bitwise weights, we have proposed a practical
implementation of the method which optimises the computational effort,
making it suitable for the large number of galaxies observed by
current/future surveys, such as BOSS, eBOSS and DESI. An important
ingredient in our modelling is given by the angular upweighting, which
allows us to minimise the variance of the estimator while not
affecting the mean.

We have provided a proof of concept of the new technique by testing it
against simulations, for different idealised observing strategies.
Besides confirming the effectiveness of the PIP weighting scheme, these
tests give us some insight into the optimal design of a survey.  Based
on our results, given a finite telescope time, it is seem more
convenient to have sparsely distributed patches with multiple
pointings rather than observing twice a single large compact area.

Although in this work we have focused on the two-point correlation function, the
reasoning behind our modelling, as well as the concept of bitwise
weights, remain valid for any $n$-pt correlation function.  We leave
to further work the interesting topic of founding a Fourier
counterpart for the PIP approach, and the practical application of
this algorithm to existing data and simulations of future data sets.

\section*{Acknowledgements}

DB and WJP acknowledge support from the European Research Council
through the Darksurvey grant 614030.  WJP also acknowledges support
from the UK Science and Technology Facilities Council grant
ST/N000668/1 and the UK Space Agency grant ST/N00180X/1.




\bibliographystyle{mnras}
\bibliography{./biblio_db}




\appendix

\section{Alternative galaxy weights}\label{sec NNvar}

We consider two more galaxy-weighting recipes designed to mitigate the
effects of missing galaxies, which share with the NN assignment the
idea of moving the weight of the discarded galaxies to the observed
ones.  First, we introduce a scheme that keeps track of the selection
history, i.e. following the idea that the individual weight of a
galaxy does not depend only on the final outcome of the selection
algorithm but also on the process that led to that outcome.
Specifically, when the MR algorithm randomly observes a galaxy out of
a (collided) pair, the weight of the discarded galaxy is assigned to
the companion, i.e. to the galaxy that actually caused the discharge.
We refer to this process as the memory dependent (MD) correction.
Second, we define a scheme in which the weight of a missing galaxy is
equally distributed (ED) to its ``direct neighbours'', i.e. the
galaxies within the distance range defined by the size of the fibre
$r_f$, the idea behind being that $r_f$ is the relevant scale for the
selection process.  Although, by construction, the MR algorithm
selects at least one galaxy per AFOF halo, it does not maximise the
number of observed objects.  This means that it is possible to have
discarded galaxies without a selected direct friend.  Therefore, to
implement the ED scheme we need to create a hierarchy in which the
discarded galaxies are classified as friend of an observed target,
friend of a friend, friend of a friend of a friend, and so on. The
weight assignment just follows this hierarchy tree, from the farthest
friends down to the observed galaxies.  As in Sec.~\ref{sec case
  study}, we get some insight on the performance of these new
weighting schemes by considering the simple AFOF structure in
Fig.~\ref{fig triplet}.  For the MD weighting scheme there are five
possible outcomes, corresponding to the rows of the following matrix
$S^{(MD)}$,
\begin{equation}
S^{(MD)} =
\begin{bmatrix}
    2 & 0 & 1 \\
    0 & 3 & 0 \\
    0 & 0 & 3 \\
    3 & 0 & 0 \\
    1 & 0 & 2 
\end{bmatrix} \quad , \quad
P^{(MD)} =
\begin{bmatrix}
    1/4  \\
    1/4  \\
    1/8 \\
    1/8  \\
    1/4 
\end{bmatrix} \ , 
\end{equation}
where $P^{(MD)}$ is the corresponding probability.
Similarly, with the ED scheme we obtain
\begin{equation}
S^{(ED)} =
\begin{bmatrix}
    3/2 & 0 & 3/2 \\
    0 & 3 & 0 \\
    0 & 0 & 3 \\
    3 & 0 & 0 
\end{bmatrix} \quad , \quad
P^{(ED)} =
\begin{bmatrix}
    1/2  \\
    1/4  \\
    1/8 \\
    1/8 
\end{bmatrix} \ . 
\end{equation}
The matrices $S^{(ED)}$ and $S^{(MD)}$ share the same properties: the
sum of the elements of each row is 3, i.e. the number of galaxy is
conserved, and the probability-weighted sum of the columns is
$\{9/8, 3/4, 9/8\}$, i.e. both the estimators are formally biased.  In
practice, we have found that, when tested against simulations, these
two corrections yield almost indistinguishable results to those
reported in Fig.~\ref{fig NN} for the NN assignment.


\section{Details on the measurements}\label{sec details}

Since our sample is a cubic box with periodic boundary conditions,
for all of the clustering measurements reported in this work we use
the natural estimator $\xi=DD/RR-1$, with analytically computed random
pair counts.  We have nonetheless checked the robustness of our results by
dropping the periodical conditions and using the standard
\citet{landy1993} estimator, with very similar results.  For the 2D
correlation function $\xi(s_\perp, s_\parallel)$ we adopt linear bins
of $1\mpcoh$ size.  The multipoles are obtained by first measuring
$\xi(s, \mu)$ and then projecting it on the Legendre polynomials
$\xi_l(s) = (2l+1)\int d\mu \ \xi(s, \mu) L_l(\mu)$.  For $\mu$ we
split the interval $[0,1]$ into $100$ linear bins.  For $s$ we adopt a
modified logarithmic binning scheme, defined by
\begin{equation}
s_i = 10^{x_0 + (i-1) \Delta x} - s_{sh} \ ,
\end{equation}
the modification being the shift term $s_{sh}$, which basically allows
us to have more control on the bin-size growth when going from small
to large scales, without loss in pair-count efficiency.  Specifically
we adopt $x_0 \approx 0.32$, $\Delta x \approx 0.038$ and
$s_{sh} = 2\mpcoh$.  For the angular pair counts $DD_a(s_\perp)$ we
use the same binning scheme with $x_0 \approx 0.34$,
$\Delta x \approx 0.080$ and $s_{sh} = 2\mpcoh$, with the only
exception of OS1 for which we adopt $x_0 \approx 0.52$,
$\Delta x \approx 0.080$ and $s_{sh} = 3\mpcoh$.

%
%
%
%
%


\bsp	
\label{lastpage}
\end{document}